\documentstyle[preprint,fancybox,epsf,amssymb,amsbsy,amstext,subeqnarray,amsfonts,aps]{revtex}
\draft
\tighten
\begin{document}

\def\dsdt{$\frac{d\sigma}{dt}$}
\def\beqn{\begin{eqnarray}}
\def\eeqn{\end{eqnarray}}
\def\barr{\begin{array}}
\def\earr{\end{array}}
\def\btab{\begin{tabular}}
\def\etab{\end{tabular}}
\def\bite{\begin{itemize}}
\def\eite{\end{itemize}}
\def\bcen{\begin{center}}
\def\ecen{\end{center}}

\def\eq{\begin{equation}}
\def\ee{\end{equation}}
\def\eqa{\begin{eqnarray}}
\def\eea{\end{eqnarray}}

\title{Fixed-t subtracted dispersion relations for Compton 
scattering off the nucleon}
\normalsize
\author{D. Drechsel, M. Gorchtein, B. Pasquini and M. Vanderhaeghen}
\address{Institut f\"ur Kernphysik, Johannes Gutenberg Universit\"at, D-55099 Mainz, Germany}
\date{\today}
\maketitle

\begin{abstract}

We present fixed-$t$ subtracted dispersion relations for Compton
scattering off the nucleon at energies $E_\gamma \leq$ 500 MeV, as a
formalism to extract the nucleon polarizabilities with a minimum of
model dependence. The subtracted dispersion integrals are mainly
saturated by $\pi N$ intermediate states in the $s$-channel 
$\gamma N \rightarrow \pi N \rightarrow \gamma N$  and $\pi \pi$
intermediate states in the $t$-channel $\gamma \gamma \rightarrow 
\pi \pi \rightarrow N \bar N$. 
For the subprocess $\gamma \gamma \rightarrow \pi \pi$, we
construct a unitarized amplitude and find a good description of the
available data. We show results for Compton scattering using the 
subtracted dispersion relations 
and display the sensitivity on the scalar 
polarizability difference $\alpha - \beta$ 
and the backward spin polarizability $\gamma_\pi$, which enter
directly as fit parameters in the present formalism. 

\end{abstract}

\section{Introduction}

Compton scattering off the nucleon is determined by 6 independent
helicity amplitudes $A_i$ ($i$ = 1,...,6), 
which are functions of two variables, e.g.
the Lorentz invariant variables $\nu$ (related to the $lab$ energy of
the incident photon) and $t$ (related to the momentum transfer to the
target). In the limit $\nu\rightarrow 0$, the general structure of
these amplitudes is governed by low energy theorems (LET) based on
Lorentz and gauge invariance. These theorems require that (I) the
leading term in the expansion in $\nu$ is determined by the global
properties of the nucleon, i.e. its charge, mass and anomalous
magnetic moment, and (II) the internal structure shows up only at
relative order $\nu^2$ and can be parametrized in terms of
polarizabilities. In this way there appear 6 polarizabilities for the
nucleon, the familiar electric and magnetic (scalar) polarizabilities
$\alpha$ and $\beta$ respectively, and 4 spin (vector) polarizabilities
$\gamma_1$ to $\gamma_4$. These polarizabilities describe the response
of the system to an external quasistatic electromagnetic field, and as
such they are fundamental structure constants of the composite system.
In particular, these polarizabilities allow one to make contact with
classical physics phenomena, the dielectric constant and the magnetic
permeability of macroscopic media as well as the Faraday effect in the
case of the spin polarizabilities.
\newline
\indent
As a consequence of LET, the differential cross section for
$\nu\rightarrow 0$ is given by the (model independent) Thomson term.
In a low-energy expansion, the electric and magnetic polarizabilities
then appear as interference between the Thomson term and the
subleading terms, i.e. as contribution of $O(\nu^2)$ in the
differential cross section, and $\alpha$ and $\beta$ can in principle
be separated by studying the angular distributions. However, it has
never been possible to isolate this term and thus to determine the
polarizabilities in a model independent way. The obvious reason is
that, for sufficiently small energies, say $\nu\leq 40$~MeV, the
structure effects are extremely small and hence the statistical errors
for the polarizabilities large. At larger energies, however, the
higher terms of the expansion, $O(\nu^4)$, become increasingly important.
Therefore, a reliable theoretical estimate of these higher terms is of
utmost importance. Moreover, at that order also the spin-dependent
polarizabilities come into the game, which has the further consequence
that a full determination of the 6 polarizabilities will require an
experimental program with polarized photons and polarized nucleons.
\newline
\indent
With the advent of high duty-factor electron accelerators and laser
backscattering techniques, new precision data have been obtained in
the 90's and more experiments are expected in the near future. In 1991
the Illinois group~\cite{Federspiel} measured differential cross
sections with tagged photons at low energy. As was to be expected, the
small counting rate and the low sensitivity to structure effects
allowed for a reduced statistical precision only. Shortly after,
Zieger et al.~\cite{Zieger} determined the cross section for photon
scattering at $\theta=180^{\circ}$ by detecting the recoil proton in a
magnetic spectrometer. In a series of experiments, the
Illinois-Saskatoon group then studied angular and energy distributions
over a wider range.  Hallin et al.~\cite{Hallin} investigated the
region from pion production threshold to the $\Delta$(1232) resonance
with a high duty-factor bremsstrahlung beam. Though the statistical
and systematical errors were small, the range of energy was clearly
outside of the low-energy expansion. The presently most accurate
values for the proton polarizabilities were derived from the work of MacGibbon
et al.~\cite{McGibbon} whose experiments were performed with tagged
photons at 70~MeV$\leq \nu \leq 100$~MeV and untagged ones at the
higher energies, and analyzed in collaboration with L'vov~\cite{lvov97}
by means of dispersion relations (in the following denoted by DR) at
constant~$t$. The final results were
\begin{eqnarray}
\alpha \;&=&\; \left(12.1 \,\pm\,0.8 \,\pm\, 0.5 \right)\,
\times\,10^{-4}\,{\mathrm fm}^3 \;, \nonumber\\  
\beta \;&=&\; \left(2.1 \,\mp\,0.8 \,\mp\,0.5 \right)\,
\times\,10^{-4}\,{\mathrm fm}^3 \;.
\label{eq1.1}
\end{eqnarray}
\newline
\indent
The physics of the $\Delta$(1232) and higher resonances has been the
objective of further recent investigations with tagged photons at
Mainz~\cite{Molinari,Peise} and with laser-backscattered photons at
Brookhaven~\cite{Tonnison}. The measured differential cross sections
and polarization asymmetries helped to discard a long-standing
problem, a unitarity violation reported in earlier experiments, and
provided useful cross-checks for the magnetic dipole and electric
quadrupole excitation of the $\Delta$ resonance. Such data were also
used to give a first prediction for the so-called backward spin
polarizability of the proton, i.e. the particular combination
$\gamma_{\pi}=\gamma_1+\gamma_2+2\gamma_4$ entering the Compton
spin-flip amplitude at $\theta=180^{\circ}$~\cite{Tonnison},
\begin{equation}
\gamma_{\pi} \;=\; -\,\left[ 27.1 \,\pm\, 2.2 ({\mathrm stat + syst})\, 
{+2.8 \atop -2.4} ({\mathrm model})\right] \,
\times\,10^{-4}\,{\mathrm fm}^4 \;.
\label{eq1.2}
\end{equation}
\newline
\indent
In 1991 Bernard et al.~\cite{Bernard} evaluated the one-loop
contributions to the polarizabilities in the framework of relativistic
chiral perturbation theory (ChPT), with the result 
$\alpha=10 \cdot \beta=12.1$, (here and in the following, 
the scalar polarizabilities are given in units of $10^{-4}$~fm$^3$). 
In order to have a systematic
chiral power counting, the calculation was then repeated in heavy
baryon ChPT~\cite{BKKM92} to $O(p^3)$, the expansion parameter being
an external momentum or the quark mass. The result explained the small
value of the magnetic polarizability, which had been difficult to
obtain in quark model calculations.  A further calculation to
$O(p^4)$ resulted in the values $\alpha=10.5\pm 2.0$ and $\beta=3.5\pm
3.6$, the errors being due to 4 counter terms entering to that order,
which were estimated by resonance saturation~\cite{BKSM}. One of these
counter terms describes the paramagnetic contribution of the
$\Delta$(1232), which is partly cancelled by large diamagnetic
contributions of pion-nucleon loops.
\newline
\indent
In view of the importance of the $\Delta$ resonance,
Hemmert, Holstein and Kambor~\cite{Holstein} proposed to include the $\Delta$
as a dynamical degree of freedom. This added a further expansion
parameter, the difference of the $\Delta$ and nucleon masses
(``$\epsilon$ expansion''). A calculation to $O(\epsilon^3)$ yielded
the results~\cite{Hemmert} (see also Ref.\cite{HHKK})
\begin{eqnarray}
\alpha & = & 12.2 + 0 + 4.2 = 16.4 \nonumber\ ,\\
\beta  & = & 1.2 + 7.2 + 0.7 = 9.1\ ,
\label{eq1.3}
\end{eqnarray}
where the 3 terms on the $rhs$ are the contributions of pion-nucleon
loops (identical to the predictions of the $O(p^3)$ calculation),
$\Delta$ pole terms, and pion-$\Delta$ loops. These predictions are
clearly at variance with the data, in particular $\alpha+\beta=25.5$
is nearly twice the rather precise value determined from DR (see
below). In an optimistic view the problem is to be cured by an
$O(\epsilon^4)$ calculation, which is likely to produce a large
diamagnetism as observed by Ref.~\cite{BKSM}. On the other hand, a
pessimist might doubt that the expansion converges sufficiently well
before the higher orders have introduced a host of unknown counter
terms.
\newline
\indent
The spin polarizabilities have been calculated in both
relativistic one-loop ChPT~\cite{BKKM92,BKKM} and heavy baryon
ChPT~\cite{HHKK}. In the latter approach the predictions are 
\begin{eqnarray}
\gamma_0 & = & 4.6-2.4-0.2+0=+2.0\ ,\nonumber \\
\gamma_{\pi} & = & 4.6+2.4-0.2-43.5=-36.7\ , 
\label{eq1.4}
\end{eqnarray}
where the spin polarizabilities are given here and in all of the following
in units of 10$^{-4}$ fm$^4$.
The 4 separate contributions on the $rhs$ of Eq.~(\ref{eq1.4}) refer to
N$\pi$-loops, $\Delta$-poles, $\Delta\pi$-loops, and the triangle
anomaly, in that order. It is obvious that the anomaly or $\pi^0$-pole
contribution is by far the most important one, and that it would
require surprisingly large higher order contributions $O(\epsilon^4)$
to increase $\gamma_{\pi}$ to the value of Ref.~\cite{Tonnison}.
Similar conclusions were reached in the framework of DR. Using the
framework of DR at $t$ = const of Ref.\cite{lvov97}, 
Ref.~\cite{DKH} obtained a value of $\gamma_{\pi}=-34.3$,
while L'vov and Nathan~\cite{LN} worked in the framework of backward
DR and predicted $\gamma_{\pi}=-39.5\pm2.4$. In the latter approach
the dispersion integral is drawn along a line $t(\nu)$ corresponding
to backward Compton scattering, i.e. on the lower boundary of the
physical $s$-channel region, which is then complemented by a path into
the physical $t$-channel region.
\newline
\indent
As we have stated before, the most quantitative analysis of the
experimental data has been provided by dispersion relations. In this
way it has been possible to reconstruct the forward non spin-flip
amplitude directly from the total photoabsorption cross section by
Baldin's sum rule~\cite{Baldin}, which yields a precise value
for the sum of the scalar polarizabilities
\begin{eqnarray}
\alpha+\beta & = & 14.2 \,\pm\, 0.5 \;\;\;
({\rm{Ref.}}\cite{Damashek}) \nonumber\\
             & = & 13.69 \,\pm\, 0.14 \;\;\; ({\rm{Ref.}}\cite{BGM})\ .
\label{eq1.5}
\end{eqnarray}
Similarly, the forward spin-flip amplitude can be evaluated by an
integral over the difference of the absorption cross sections in
states with helicity 3/2 and 1/2,
\begin{eqnarray}
  \gamma_0 = \gamma_1-\gamma_2-2\gamma_4 & = & -1.34\; \; 
  ({\rm{Ref.}}\cite{Sandorfi}) \nonumber \\ 
   & = & -0.6\; \; ({\rm{Ref.}}\cite{DKH})\ .
\label{eq1.6}
\end{eqnarray}
While these predictions rely on pion photoproduction multipoles, the
helicity cross sections have now been directly determined by
scattering photons with circular polarizations on polarized
protons~\cite{Ahrens}.
\newline
\indent
In the case of forward Compton scattering the momentum transfer and,
hence, the Mandelstam variable $t$ vanishes. In that sense the above
sum rules of Eqs. (\ref{eq1.5}, \ref{eq1.6}) 
are derived from DR at $t=0$. At finite angles, however, the
analysis requires DR at $t={\rm{const.}}\leq0$, in a range of values
between 0 (forward scattering) and the largest negative value of $t$
($t = t_{{\rm{max}}}$), determined by the
largest scattering angle at the highest photon energy. As mentioned
above, the most quantitative and detailed such analysis has been
performed by L'vov and collaborators~\cite{lvov97,lvov98} in the framework of
unsubtracted DR at $t=$ const. Unfortunately, not all of the
dispersion integrals converge, as can be inferred from Regge theory.
The reason for the divergence of the integrals is essentially given by
fixed poles in the $t$ channel, notably the exchange of the neutral
pion and of a somewhat fictitious $\sigma$ meson with a mass of about
600~MeV and a large width, which models the two-pion continuum with
the quantum numbers $I=J=0$. In a more formal view, the dispersion
integral is performed along the real axis in the range
$-\nu_{max}\leq\nu\leq+\nu_{max}$, with $\nu_{max}\approx$ 1.5~GeV, and then
closed by a semi-circle with radius $\nu_{max}$ in the upper half of the
complex $\nu$-plane. The contribution of the semi-circle is then
identified with the asymptotic contribution described by $t$-channel
poles. This introduces unknown vertex functions and the mass 
of the ``$\sigma$ meson'', which have to be determined from the
experiment. Moreover, the analysis depends appreciably on the choice
of $\nu_{max}$, and there are substantial contributions of intermediate
states beyond the relatively well-known pion-nucleon continuum. These
higher states include multipion, $\eta$- and $\rho$-meson production,
$\Delta \pi$-loops and nonresonant s-wave background. The physics
behind these effects is certainly worthwhile studying, and there can
be no doubt that within the next years we shall learn more about them
by detailed coincidence studies of multipion and heavier meson
production, but also directly from a careful analysis of Compton
scattering at the higher energies~\cite{lvov97}. However, the quest for
the polarizabilities as fundamental structure constants should not be
burdened by too many open questions and phenomenological models.
\newline
\indent
In view of the problems of unsubtracted DR, we propose to analyse
Compton scattering in the framework of subtracted DR at constant $t$,
with the eventual goal to determine the 6
polarizabilities with the least possible model dependence. We choose
to subtract the 6 Compton amplitudes $A_i(\nu, t)$ at the unphysical
value $\nu=0$, i.e. write subtracted DR for $A_i(\nu, t)-A_i(0, t)$ at
constant $t$. As we shall show in the following, these subtracted DR
converge nicely and are quite well saturated already at $\nu\approx
400-550$~MeV, i.e. essentially by one-pion production. Clearly the
price to pay are 6 new functions $A_i(0, t)$, which have to be
determined by another set of dispersion relations, at $\nu$=const=0 and
by use of information obtained from the $t$-channel reaction
$\gamma\gamma\rightarrow$ anything.
\newline
\indent
In order to reduce the dependence on the higher intermediate states in
the $t$-channel, we subtract again, i.e. write DR for $A_i(0,
t)-A_i(0, 0)$, the subtraction constants $A_i(0, 0)$ being linear
combinations of the 6 polarizabilities. Since 4 of these subtraction
constants can be calculated from unsubtracted DR at $t$=const, only 2
parameters have to be fixed by a fit to low energy Compton
scattering, the combinations $\alpha-\beta$ and
$\gamma_{\pi}=\gamma_1+\gamma_2+2\gamma_4$ describing the backward
non spin-flip and spin-flip amplitudes, respectively.
\newline
\indent
In a somewhat similar approach, Holstein and Nathan~\cite{HN} combined
$s$- and $t$-channel information to predict the backward scalar
polarizability $\alpha-\beta$. Using unsubtracted backward DR they
obtained, from the integration along the lower boundary of the
$s-$channel region, the result $(\alpha-\beta)^s=-6\pm3$, and from the
$t$-channel region a contribution of $(\alpha-\beta)^t\approx9$. The
sum of these two contributions, $\alpha-\beta\approx3\pm3$, is at
variance with the presently accepted experimental (global average) value,
$\alpha-\beta=10.0\pm1.5\pm0.9$ of Ref.~\cite{McGibbon}. The
difficulty to predict this observable is due to the bad
convergence of the integrals in both the $s-$ and the $t$-channel
regions. As may be seen from Fig.~3 of Ref.~\cite{HN}, the
$t$-channel integral obtains quite sizeable contributions from
$10 \, m_{\pi}^2<t<40 \, m_{\pi}^2$, in which region the integrand changes its
sign. Independent of the numerical analysis, the authors find an
extremely interesting relationship connecting the polarizabilities of
pions and nucleons. This connection comes via the $t$-channel integral
for the nucleon and the low-energy expansion of the $s$-wave
$\gamma\gamma\rightarrow\pi\pi$ amplitude, and results in
$\delta\alpha=-\delta\beta=0.5\alpha^{\pi}\approx 1.4$, which is the
contribution of the pion polarizability $\alpha^{\pi}$ to the
nucleon's electric and magnetic polarizabilities. In concluding this
discussion we point out the difference of our present approach and the
calculation of Ref.~\cite{HN}. We do not intend to predict
$\alpha-\beta$. Instead we want to develop a dispersion description of
Compton scattering allowing for a derivation of $\alpha-\beta$ with a
minimum of model dependence. For this purpose we use a scheme of
subtracted DR whose subtraction constants are linear
combinations of polarizabilities, which have to be determined by a fit
to the Compton data.
\newline
\indent
In section~\ref{disp} we shall give a general introduction to subtracted DR.
This technique is then applied to the cases of DR at $t$=const
($s$-channel dispersion integral) and DR at $\nu=0$ ($t$-channel
dispersion integral) in sections~\ref{schannel} 
and \ref{tchannel}, respectively. Our results
are compared to the existing low-energy Compton data in 
section~\ref{results}, and our conclusions are drawn in 
section~\ref{conclusions}.

\section{Fixed-$t$ subtracted dispersion relations}
\label{disp}

Assuming invariance under parity, charge conjugation and time reversal
symmetry, the general amplitude for Compton scattering can be
expressed in terms of six independent structure functions $A_i(\nu,
t)$, $i=1,...,6$. These structure functions depend on two Lorentz
invariant variables, e.g. $\nu$ and $t$ as defined in the following.
Denoting the momenta of the initial state photon and proton by $q$ and
$p$ respectively, and with corresponding final state momenta
 $q'$ and $p'$, the familiar Mandelstam variables are
\begin{equation}
s=(q+p)^2\ ,\ \ t=(q-q')^2\ ,\ \ u=(q-p')^2\ .
\label{eq2.1}
\end{equation}
These variables fulfill the constraint
\begin{equation}
s+t+u=2M^2\ .
\label{eq2.2}
\end{equation}
The variable $\nu$ is defined by,
\begin{equation}
\nu=\frac{s-u}{4M}=E_{\gamma}+\frac{t}{4M}\ ,
\label{eq2.3}
\end{equation}
where $E_{\gamma}$ is the photon energy in the $lab$ frame and $M$ the
nucleon mass. The Mandelstam plane is shown in Fig.~1, and the
boundaries of the physical and spectral regions are discussed in
Appendix~A.
\newline
\indent
The invariant amplitudes $A_i$ are free of kinematical singularities
and constraints, and because of the crossing symmetry they satisfy the
relation $A_i(\nu, t)=A_i(-\nu, t)$. Assuming further analyticity and
an appropriate high-energy behavior, the amplitudes $A_i$ fulfill 
unsubtracted dispersion relations at fixed $t$,
\begin{equation}
{\mathrm Re} A_i(\nu, t) \;=\; A_i^B(\nu, t) \;+\;
{2 \over \pi} \; {\mathcal P} \int_{\nu_{thr}}^{+ \infty} d\nu' \; 
{{\nu' \; {\mathrm Im}_s A_i(\nu',t)} \over {\nu'^2 - \nu^2}}\;,
\label{eq:unsub} 
\end{equation}
where $A_i^B$ are the Born (nucleon pole) contributions, ${\mathrm
  Im}_s A_i$ the discontinuities across the $s$-channel cuts of the
Compton process and $\nu_{thr} = m_\pi + (m_\pi^2 + t/2)/(2 M)$. 
However, such unsubtracted dispersion relations
require that at high energies ($\nu \rightarrow \infty$) the
amplitudes ${\mathrm Im}_s A_i(\nu,t)$ drop fast enough so that the
integral of Eq.~(\ref{eq:unsub}) is convergent and the contribution from
the semi-circle at infinity can be neglected. For real Compton
scattering, Regge theory predicts the following high-energy behavior
for $\nu \rightarrow \infty$ and fixed $t$~\cite{lvov97}:
\begin{eqnarray}
&&A_{1,2} \sim \nu^{\alpha(t)} \;,\nonumber\\
&&A_{3,5,6} \sim \nu^{\alpha(t) - 2} \;,\hspace{1cm} 
A_{4} \sim \nu^{\alpha(t) - 3} \;,
\end{eqnarray}
where $\alpha(t) \lesssim 1$ is the Regge trajectory. In particular we
note that the Regge trajectory with the highest intercept, i.e.
$\alpha(0) \approx 1.08$, corresponds to soft pomeron exchange. Due to
this high energy behavior, the unsubtracted dispersion integral of
Eq.~(\ref{eq:unsub}) diverges for the amplitudes $A_1$ and $A_2$. In order
to obtain useful results for these two amplitudes, L'vov et
al.~\cite{lvov97} proposed to close the contour of the integral in
Eq.~(\ref{eq:unsub}) by a semi-circle of finite radius $\nu_{max}$
(instead of the usually assumed infinite radius!) in the complex
plane, i.e.  the real parts of $A_1$ and $A_2$ are calculated from the
decomposition
\begin{equation}
{\mathrm Re} A_i(\nu, t) \;=\; A_i^B(\nu, t) \;+\;
A_i^{int}(\nu, t) \;+\; A_i^{as}(\nu, t) \;,
\label{eq:aintas}
\end{equation}
with $A_i^{int}$ the $s$-channel integral from pion
threshold $\nu_{thr}$ to a finite upper limit $\nu_{max}$,
\begin{equation}
A_i^{int}(\nu, t) \;=\; 
{2 \over \pi} \; {\mathcal P} \int_{\nu_{thr}}^{\nu_{max}} d\nu' \; 
{{\nu' \; {\mathrm Im}_s A_i(\nu',t)} \over {\nu'^2 - \nu^2}}\;,
\label{eq13}
\end{equation}
and an `asymptotic contribution' $A_i^{as}$ representing the
contribution along the finite semi-circle of radius $\nu_{max}$ in the
complex plane. In the actual calculations, the $s$-channel integral is
typically evaluated up to a maximum photon energy $E_\gamma =
\nu_{max}(t) - t/(4 M) \approx 1.5$~GeV, for which the imaginary parts
of the amplitudes can be expressed through unitarity by the meson
photoproduction amplitudes (mainly 1$\pi$ and 2$\pi$ photoproduction)
taken from experiment.  All contributions from higher energies are
then absorbed in the asymptotic term, which is replaced by a finite
number of energy independent poles in the $t$-channel. In particular
the asymptotic part of $A_1$ is parametrized in Ref.\cite{lvov97} 
by the exchange of a scalar particle in the $t$-channel, i.e. an effective
``$\sigma$ meson'',
\begin{equation}
A_1^{as}(\nu, t) \approx A_1^{\sigma}(t) \;=\;
{{F_{\sigma \gamma \gamma} \; g_{\sigma NN}} 
\over {t - m_\sigma^2}} \;,  
\end{equation}
where $m_\sigma$ is the ``$\sigma$ meson'' mass, and 
$g_{\sigma NN}$ and $F_{\sigma \gamma \gamma}$ are the couplings of the
``$\sigma$ meson'' to the nucleons and photons respectively.
The asymptotic part of $A_2$ is parametrized by the $\pi^0$
$t$-channel pole. 
\newline
\indent
This procedure is relatively save for $A_2$ because of the dominance
of the $\pi^0$ pole or triangle anomaly, which is well established
both experimentally and on general grounds as Wess-Zumino-Witten term.
However, it introduces a considerable model-dependence in the case of
$A_1$. Though ``$\sigma$ mesons'' have been repeatedly reported in the
past, their properties were never clearly established. Therefore, this
particle should be interpreted as a parametrization of the $I=J=0$
part of the two-pion spectrum, which shows up differently in different
experiments and hence has been reported with varying masses and
widths.
\newline
\indent
It is therefore the aim of our present contribution to avoid the
convergence problem of unsubtracted DR and the
phenomenology necessary to determine the asymptotic contribution. The
alternative we shall pursue in the following is to consider
DR at fixed $t$ that are once subtracted at $\nu=0$,
\begin{equation}
{\mathrm Re} A_i(\nu, t) \;=\; A_i^B(\nu, t) \;+\;
\left[ A_i(0, t) - A_i^B(0, t) \right]
\;+\;{2 \over \pi} \;\nu^2\; {\mathcal P} \int_{\nu_{thr}}^{+ \infty} d\nu' \; 
{{\; {\mathrm Im}_s A_i(\nu',t)} \over {\nu' \; (\nu'^2 - \nu^2)}}\;.
\label{eq:sub} 
\end{equation}
These subtracted DR should converge for all six invariant amplitudes
due to the two additional powers of $\nu'$ in the denominator, and they are
essentially saturated by the $\pi N$ intermediate states as will be
shown in section \ref{schannel}. In other
words, the lesser known contributions of two and more pions 
as well as higher continua are small and may be treated reliably
by simple models.
\newline
\indent
The price to pay for this alternative is the appearance of the
subtraction functions $A_i(\nu=0, t)$, which have to be determined at
some small (negative) value of $t$. We do this by setting up a
once-subtracted DR, this time in the variable $t$,
\begin{eqnarray}
A_i(0, t) \;-\; A_i^B(0, t) &=&
\left[ A_i(0, 0) \;-\; A_i^B(0, 0) \right] 
\;+\;
\left[ A_i^{t-pole}(0, t) \;-\; A_i^{t-pole}(0, 0) \right] \nonumber\\ 
&+&\;{t \over \pi} \; \int_{(2 m_\pi)^2}^{+ \infty} dt' \;
{{{\mathrm Im}_t A_i(0,t')} \over {t' \; (t' - t)}}
\;+\;{t \over \pi} \; \int_{- \infty}^{a} dt' \;
{{{\mathrm Im}_t A_i(0,t')} \over {t' \; (t' - t)}} \;,
\label{eq:subt} 
\end{eqnarray}
where $A_i^{t-pole}(0, t)$ represents the contribution of poles in the
$t$-channel, in particular of the $\pi^0$ pole in the case of $A_2$, 
which is given by
\begin{equation} 
A_2^{\pi^0}(0, t) \;=\; {{F_{\pi^0 \gamma \gamma} \; g_{\pi NN}} 
\over {t - m_\pi^2}} \;.
\label{eq:piopole}
\end{equation}
The coupling $F_{\pi^0 \gamma \gamma}$ is determined through the $\pi^0
\rightarrow \gamma \gamma$ decay as 
\begin{equation}
\Gamma\left( \pi^0 \rightarrow \gamma \gamma\right) \;=\; 
{1 \over {64 \, \pi} } \, m_{\pi^0}^3 \, F_{\pi^0 \gamma \gamma}^2 \;.
\end{equation}
Using $\Gamma\left( \pi^0 \rightarrow \gamma \gamma\right)$ = 7.74 eV
\cite{PDG98}, one obtains 
$F_{\pi^0 \gamma \gamma}$ = -0.0252 GeV$^{-1}$,
where the sign is in accordance with the $\pi^0 \gamma \gamma$
coupling in the chiral limit, given by the Wess-Zumino-Witten 
effective chiral Lagrangian. 
The $\pi NN$ coupling is taken from Ref.~\cite{arndtpin98} : $g_{\pi
  NN}^2/4 \pi$ = 13.72. This yields then for the product of the couplings in
Eq.~(\ref{eq:piopole}) : $F_{\pi^0 \gamma \gamma} \, g_{\pi NN}
\approx$ -0.331 GeV$^{-1}$.  
\newline
\indent
The imaginary part in the integral from $4 m_\pi^2 \rightarrow \infty$
in Eq.~(\ref{eq:subt}) 
is saturated by the possible intermediate states for the $t$-channel
process (see Fig.~\ref{fig:tunit}), 
which lead to cuts along the positive $t$-axis. 
For values of $t$ below the $K \bar K$
threshold, the $t$-channel discontinuity is dominated by $\pi \pi$
intermediate states.  The second integral in Eq.~(\ref{eq:subt})
extends from $-\infty$ to $a$, where 
$a \,=\, -4\, (m_{\pi}^2 + 2 M m_\pi) \approx - 1.1$ GeV$^2$ is the
boundary of the $su$ spectral region for $\nu = 0$ (see Appendix
\ref{app:spectralcompton} for a detailed discussion). As we are
interested in evaluating Eq.~(\ref{eq:subt}) for small (negative)
values of $t$ ($|t|\ll |a|$), the integral from $- \infty$ to
$a$ will be highly suppressed by the denominator of the subtracted DR,
and can therefore be neglected.  Consequently, we shall saturate the
subtracted dispersion integrals of Eq.~(\ref{eq:subt}) by the
contribution of $\pi\pi$ intermediate states, which turns out to be a
good approximation for small $t$. We will show the convergence of
the $t$-channel dispersion integral in section \ref{tchannel} 
and thus verify the quality of the approximation.
\newline
\indent
The $t$-dependence of the subtraction functions $A_i(0, t)$ is now
determined, and only the subtraction constants $A_i(0, 0)$ remain to
be fixed. We note that the quantities
\begin{equation}
a_i = A_i(0, 0) \;-\; A_i^B(0, 0)
\label{eq.a_i}
\end{equation}
are directly related to the polarizabilities, which can then be
obtained from a fit to the Compton scattering data. 
For the spin-independent (scalar) polarizabilities $\alpha$ and
$\beta$, one finds the two combinations
\begin{eqnarray}
\label{eq:alphaplusbeta}
\alpha + \beta \;&=&\; - {1 \over {2 \pi}} \; (a_3 \;+\; a_6)\;, \\
\alpha - \beta \;&=&\; - {1 \over {2 \pi}} \; a_1\;, 
\label{eq:nospinpol0}
\end{eqnarray}
which can be determined from forward and backward scattering
respectively. Furthermore, the forward combination $\alpha + \beta$ is
related to the total absorption spectrum through Baldin's sum rule 
\cite{Baldin},
\begin{eqnarray}
(\alpha + \beta)_N \;&=&\
{1 \over {2 \pi^2}} \; \int_{\nu_{thr}}^{\infty} d\nu' \; 
{{\sigma(\gamma N \rightarrow X)} \over {\nu'^2}}\;.
\label{eq22}
\end{eqnarray}
The 4 spin dependent polarizabilities $\gamma_1$ to 
$\gamma_4$ of Ragusa~\cite{ragusa} are defined by
\begin{eqnarray}
\gamma_0 \;&\equiv&\; \gamma_1 - \gamma_2 - 2 \gamma_4 \;=\;
{1 \over {2 \pi} M} \; a_4\;, \label{eq23} \\
\gamma_{13} \;&\equiv&\; \gamma_1 + 2 \gamma_3 \;=\;
-\;{1 \over {4 \pi} M} \; (a_5 \;+\; a_6)\;, \label{eq24} \\
\gamma_{14} \;&\equiv&\; \gamma_1 - 2 \gamma_4 \;=\;
{1 \over {4 \pi} M} \; (2 \, a_4 \;+\; a_5 \;-\; a_6)\;, \label{eq25} \\
\gamma_\pi \;&\equiv&\; \gamma_1 + \gamma_2 + 2 \gamma_4 \;=\;
-\;{1 \over {2 \pi} M} \; (a_2 \;+\; a_5)\;,
\label{eq:spinpol0}
\end{eqnarray}
where $\gamma_0$ and $\gamma_\pi$ are the spin (vector)
polarizabilities in the forward and backward directions respectively.
Since the $\pi^0$ pole contributes to $A_2$ only, the combinations
$\gamma_0$, $\gamma_{13}$ and $\gamma_{14}$ of Eqs.~(\ref{eq23})-(\ref{eq25})
are independent of this pole term~\cite{DKH}, and only the backward
spin polarizability $\gamma_\pi$ is affected by the anomaly. 
\newline
\indent
Although all 6 subtraction constants $a_1$ to $ a_6$ of
Eq.~(\ref{eq.a_i}) could be used as fit parameters, we shall restrict
the fit to the parameters $a_1$ and $a_2$, or equivalently to
$\alpha - \beta$ and $\gamma_\pi$.  The subtraction constants $a_4,
a_5$ and $a_6$ will be calculated through an unsubtracted sum rule,
as derived from Eq.~(\ref{eq:unsub}),
\begin{equation}
a_{4, 5, 6} \;=\; {2 \over \pi} \; \int_{\nu_{thr}}^{+ \infty} d\nu' \; 
{{ {\mathrm Im}_s A_{4, 5, 6}(\nu',t = 0)} \over {\nu'}}\;.
\label{eq:a4a5a6}
\end{equation}
The remaining subtraction constant $a_3$, 
which is related to $\alpha + \beta$ through Eq.~(\ref{eq:alphaplusbeta}), 
will be fixed through Baldin's sum rule, Eq.~(\ref{eq22}), using the
value obtained in Ref.~\cite{BGM} : $\alpha + \beta = 13.69$.

\section{s-channel dispersion integral}
\label{schannel}

In this section we describe the calculation of the $s$-channel
contributions, which enter in the once-subtracted 
dispersion integral of Eq.~(\ref{eq:sub}) and in the calculation of
subtraction constants $a_4,\, a_5$ and $a_6$ through Eq.~(\ref{eq:a4a5a6}). 
The imaginary part of the Compton amplitude due to the $s$-channel
cuts is determined from the scattering amplitudes of
photoproduction on the nucleon by the unitarity relation
\begin{equation}
\label{s-unit}
2\,\mbox{Im}_s\,T_{fi}=
\sum_X (2\pi)^4 \delta^4(P_X-P_i)T^{\dagger}_{X f }\,T_{X i} \;,
\end{equation}
where the sum runs over all possible states that can be formed in the
photon-nucleon reaction.  Due to the energy denominator 
$1/\nu'(\nu'^2-\nu^2)$ in the subtracted dispersion integrals, 
the most important contribution is from the $\pi
N$ intermediate states, while mechanisms involving more pions or
heavier mesons in the intermediate states 
are largely suppressed.  In our calculation, we evaluate the $\pi N$
contribution using the multipole amplitudes from the analysis of
Hanstein, Drechsel and Tiator~\cite{HDT} at energies $E_\gamma\le
500$~MeV and at the higher energies we take as input the SAID
multipoles (SP98K solution)~\cite{said}.  The expansion of
$\mbox{Im}_s A_i$ into this set of multipoles is truncated at a
maximum angular momentum $j_{max}=l\pm 1/2=7/2,$ with the exception of
the lower energy range ($E_\gamma\le 400$~MeV) where we use $j_{{\rm
    max}}=3/2$.  The higher partial waves with $j\ge j_{max}+1$ are
evaluated analytically in the one-pion exchange (OPE) approximation.
The relevant formulas to implement the calculation are reported in
Appendix B and C of Ref.\cite{lvov97}.
\newline
\indent
The multipion intermediate states are approximated by the inelastic
decay channels of the $\pi N$ resonances.  In the spirit of
Ref.~\cite{lvov97} and the more recent work of Ref.~\cite{kamalov}, we
assume that this inelastic contribution follows the helicity structure
of the one-pion photoproduction amplitudes. In this approximation, we
first calculate the resonant part of the pion photoproduction multipoles
using the Breit-Wigner parametrization of Ref.~\cite{arndt}, which is then
scaled by a suitable factor to include the inelastic decays of the
resonances.  The resulting contribution to $\mbox{Im}_s\, A_i$ is
\begin{eqnarray}
\left[\mbox{Im}_s\, A_i\right]^{(N^*\rightarrow \pi\pi N,\eta N,...)}=R
\left[\mbox{Im}_s\, A_i\right]^{(N^*\rightarrow \pi N)},
\end{eqnarray}
with the ratio $R$ given by  
\begin{eqnarray}
\label{tp-scale}
R=\frac{1-B_\pi}{B_\pi}\frac{\bar\Gamma_{{\rm inel}}(W)}{\bar\Gamma_\pi(W)}.
\end{eqnarray}
In Eq.~(\ref{tp-scale}), $B_\pi$ is the single-pion branching ratio of the
resonance $N^*$ and $\bar\Gamma_\pi(W)$ the energy-dependent pionic
width~\cite{arndt}, while the inelastic width $\bar\Gamma_{{\rm
    inel}}(W)$ of the decays $N^*\rightarrow(\pi\pi N,\,\eta
N,\,\pi\pi\pi N,..)$ is parametrized as in Ref.\cite{lvov97} in order
to provide the correct threshold behavior for the resonant two pion
contribution.
\newline
\indent
The $\pi N$ channel consistently reproduces the measured
photoabsorption cross section in the energy range $E_\gamma\le
500$~MeV, while at the higher energies nonresonant mechanisms should
be included in addition to the resonant mechanism 
to fully describe the multipion channels.  In
Ref.\cite{lvov97}, the non-resonant contribution 
to the two-pion photoproduction channel was approximately taken into
account by calculating the OPE diagram 
of the $\gamma N\rightarrow \pi\Delta$ reaction. 
The difference between the data and the model for two-pion photoproduction 
consisting of resonant mechanisms plus the OPE
diagram for the nonresonant mechanism, was then fitted in
Ref.\cite{lvov97} and attributed to a phenomenological, 
non-resonant $\gamma N\rightarrow \pi\Delta$ s-wave correction term. 
\newline
\indent
A more detailed description of the $\pi \pi N$ channel is clearly
worthwile to be undertaken, especially in view of the new 
two-pion photoproduction data (both unpolarized and polarized) that will be
available from MAMI and JLab (CLAS) in the near future. 
However, for the extraction of the polarizabilities, the strategy
followed in this paper is to minimize sensitivity and hence model
uncertainty to these higher channels. 
\newline
\indent
We show in Fig.~\ref{fig:a1a2conv} 
that in the subtracted DR, the
sensitivity to the multipion channels is indeed very small. 
For the unsubtracted DR, on the other hand, 
the influence of the multipion channels 
amounts to about 30 \% of the amplitude $A_2$. 
We furthermore note from Fig.~\ref{fig:a1a2conv} that the subtracted DR
are essentially saturated at $\nu~\approx$ 0.4 GeV and 
that they only receive a negligible contribution from multipion
channels. The importance of the multipion channels is even weaker in 
the case of the amplitudes $A_3$ to $A_6$.  
\newline
\indent
In Table I and II, we show our predictions for the
dispersion integral of the spin polarizabilities of the proton and
neutron, respectively.  We list the separate contribution of the $\pi
N$ channel, HDT($1\pi$), and the total result, HDT, which includes
the inelastic resonance channels.  The last column shows the values of
the dispersion calculation of Ref.\cite{lvov98}, which is based on the
one-pion multipoles of the SAID-SP97K solution and the model for
double-pion production mentioned above. The small differences between
the one-pion multipoles of SAID-SP97K and SAID-SP98K at the higher
energies are practically negligible for the spin polarizabilities,
while the results are very sensitive to the differences between the
HDT and SAID analyses.  As discussed in Ref.~\cite{DK},
this fact is mainly due to a different behaviour of the $E_{0+}$
partial wave near threshold, giving rise to substantial effects in the
case of the forward spin polarizability. While the one-pion
contribution from SAID-SP98K is $\gamma^{p}_0=-1.26$ and
$\gamma^{n}_0=-0.03 $, we obtain $\gamma^{p}_0=-0.75$ and
$\gamma^{n}_0=-0.06$ with the HDT multipoles for $E_\gamma\le 500$~MeV.

\section{t-channel dispersion integral}
\label{tchannel}

We next evaluate the $t$-channel dispersion integral in
Eq.~(\ref{eq:subt}) from $4 m_\pi^2$ to $\infty$. 
The kinematics of the $t$-channel reaction $\gamma \gamma \rightarrow
N \bar N$ is shown in Fig.~\ref{fig:tkinematics}.
The subtracted dispersion integral is essentially saturated by 
the imaginary part of the $t$-channel amplitude 
$\gamma \gamma \rightarrow N \bar N$ due to 
$\pi\pi$ intermediate states. 
To calculate this contribution, we 
have to construct the amplitudes $\gamma \gamma \rightarrow \pi \pi$
and $\pi \pi \rightarrow N \bar N$.
\newline
\indent
We start with the isospin and helicity structure 
of the $\gamma \gamma \rightarrow \pi \pi$ amplitude, denoted by $F$. 
Because of the Bose symmetry of the $\gamma\gamma$
state, only the even isospin values $I$ = 0 and 2 are possible.  We
can express the charged ($\gamma \gamma \rightarrow \pi^+ \pi^-$) 
and neutral ($\gamma \gamma \rightarrow \pi^0 \pi^0$) amplitudes 
in terms of those with good isospin by
\begin{eqnarray}
{F^{(\pi^+ \pi^-)}}\;=\;
\sqrt{2 \over 3}\;{F^{I=0}}\;+\;\sqrt{1 \over 3}\;{F^{I=2}}
{\mathrm \;\;\;(charged\;pions) \;,}\nonumber\\
{F^{(\pi^0 \pi^0)}}\;=\;
-\sqrt{1 \over 3}\;{F^{I=0}}\;+\;\sqrt{2 \over 3}\;{F^{I=2}}
{\mathrm \;\;\;(neutral\;pions) \;.}
\label{eq:isogagapipi}
\end{eqnarray}
The reaction $\gamma \gamma \rightarrow \pi \pi$ 
has two independent helicity amplitudes
${F}_{\Lambda_\gamma}(t, \theta_{\pi \pi})$, 
where $\Lambda_\gamma \equiv 
\lambda'_\gamma - \lambda_\gamma$, 
being the difference of the final photon helicity ($\lambda'_\gamma$)
and the initial photon helicity ($\lambda_\gamma$), 
takes on the values 0 or 2,
depending upon whether
the photons have the same ($\Lambda_\gamma$ = 0) or
opposite ($\Lambda_\gamma$ = 2) helicities. 
The $\gamma \gamma \rightarrow \pi \pi$ 
helicity amplitudes depend upon the c.m. energy squared $t$, and the
pion c.m. scattering angle $\theta_{\pi \pi}$.
In terms of the helicity amplitudes $F_{\Lambda_\gamma}$, the 
$\gamma \gamma \rightarrow \pi \pi$ differential c.m. cross section is
given by 
\begin{eqnarray}
{\left(\frac{d\sigma}{d\cos\theta_{\pi \pi}}\right)_{\rm c.m.}}\;=\;
\frac{\beta}{64 \, \pi \,t} \;
\left\{|{{F}_{\Lambda_\gamma=0}}(t,\theta_{\pi \pi})|^2\;+\;
|{{F}_{\Lambda_\gamma=2}}(t,\theta_{\pi \pi})|^2\right\} \;,
\label{eq:crossgagapipi}
\end{eqnarray}
with $\beta = \sqrt{1 - 4 m_\pi^2/t}$ the pion velocity. 
In Appendix \ref{app:tchannel}, we give the partial wave expansion of
the $\gamma \gamma \rightarrow \pi \pi$ helicity amplitudes
$F^I_{\Lambda_\gamma}(t, \theta_{\pi \pi})$ for a state of isospin $I$, 
and thus define the partial wave amplitudes 
$F^I_{J \,\Lambda_\gamma}(t)$ 
(see Eqs.~(\ref{eq:partialgagapipi}) and (\ref{eq:deffgagapipi})), 
where $J$ can only take on even values.
\newline
\indent
To construct the helicity amplitudes $F_{\Lambda_\gamma}$ for the
process $\gamma\gamma\,\rightarrow\,\pi\pi$, 
we first evaluate the Born graphs as shown in Fig.~\ref{fig:born}.
These graphs only contribute to the charged channel
$\gamma\gamma\,\rightarrow\,\pi^+\pi^-$.  The Born
contributions to the helicity amplitudes 
${F}^{(\pi^+ \pi^-)}_{\Lambda_\gamma}$ are denoted as  
${B}_{\Lambda_\gamma}$ and given by
\begin{eqnarray}
{B_{\Lambda_\gamma = 0}}(t,\theta_{\pi \pi})
&&\;=\;\left( 2 e^2\right)\,\frac{1\,-\,\beta^2}
{1\,-\,{\beta^2}{\cos^2}\theta_{\pi \pi}} \;,\nonumber\\
{B_{\Lambda_\gamma = 2}}(t,\theta_{\pi \pi})
&&\;=\;\left( 2 e^2\right)\,\frac{{\beta^2}{\sin^2}\theta_{\pi \pi}}
{1\,-\,{\beta^2}{\cos^2}\theta_{\pi \pi}} \;.
\label{eq:gagapipibornhel}
\end{eqnarray}
The partial wave expansion of the Born terms 
$B_{J \, \Lambda_\gamma}(t)$ is discussed in  
Appendix \ref{app:tchannel} (Eq.~(\ref{eq:gagapipiborn})).
As the Born amplitudes are only non-zero for the charged pion channel,
the two isospin amplitudes of Eq.(\ref{eq:isogagapipi}) are related by 
\begin{equation}
{B^{I=0}_{J\Lambda_\gamma}}\;=\;\sqrt{2 \over 3}\;{B_{J\Lambda_\gamma}}\,,
\hspace{2cm}
{B^{I=2}_{J\Lambda_\gamma}}\;=\;\sqrt{1 \over 3}\;{B_{J\Lambda_\gamma}}\,.
\end{equation}
\newline
\indent
We now construct the unitarized amplitudes $F^I_{J \Lambda_\gamma}(t)$,
starting from the Born amplitudes $B^I_{J \Lambda_\gamma}(t)$ and
following the method outlined in Refs.~\cite{Morgan88,Pennington95}. We first
note that the low energy theorem requires for each partial wave that
\begin{equation}
\frac{F^{I}_{J\Lambda_\gamma}}{B^{I}_{J\Lambda_\gamma}}\,\rightarrow\,1\,,\;
{\mathrm as}\;\;t\,\rightarrow\,0\,.
\end{equation}
Next, the invariant amplitude for the process
$\gamma\gamma\,\rightarrow\,\pi\pi$ is assumed to have Mandelstam
analyticity. Each partial wave then has a right-hand cut from
$t\,=\,4m^2_\pi$ to $+\infty$ and a left-hand cut from $t\,=\,-\infty$
to $0$. Though the Born amplitude is real for all values of $t$, its
partial waves are complex below $t=0$. The
partial waves of the full amplitude have no other sources of
complexity in this region, and so we can write a 
DR for the difference of the full and the Born amplitudes,
\begin{eqnarray}
\frac{F^{I}_{J\Lambda_\gamma}(t)\,-\,B^{I}_{J\Lambda_\gamma}(t)}
{t(t\,-\,4{m^2_\pi})^{J\over2}}\;=\;\frac{1}{\pi}\;
{\int^\infty_{4m^2_\pi}}\;dt'\;
\frac{{\mathrm Im}F^{I}_{J\Lambda_\gamma}(t')}
{t'(t'\,-\,4{m^2_\pi})^{J\over2}(t'-t)}\,,
\label{eq:drgagapipi}
\end{eqnarray}
with an additional factor of $({t(t\,-\,4{m^2_\pi})^{J\over2}})^{-1}$
providing the right asymptotics for the convergence of the integral.
The next step is to evaluate the imaginary part of
the amplitude in Eq.~(\ref{eq:drgagapipi}). 
To do this, we exploit the unitarity condition
\begin{eqnarray}
{\mathrm Im}\,{{F}^I_{J \Lambda_\gamma}}
(\gamma\gamma\,\rightarrow\,\pi\pi)\;=\;
{\sum_n}\;{\rho_n}\;
{{F}^{I \ast}_{J \Lambda_\gamma}}(\gamma\gamma\,\rightarrow\,n{)}\;
{{\cal{I}}^I_{J}}(n\,\rightarrow\,\pi\pi)\,,
\label{eq.Im}
\end{eqnarray}
where $\rho_n$ are the appropriate kinematical and isospin factors for
the intermediate channels $n$, and ${\cal I}(n \rightarrow \pi\pi)$ is
a hadronic amplitude. 
Below the next inelastic threshold, it follows from unitarity that the phase
${\phi^{I\;(\gamma\gamma\,\rightarrow\,\pi\pi)}_{J}}$ of
each partial wave $F^I_{J \, \Lambda_\gamma}$ 
is equal to the phase $\delta^{I\,J}_{\pi\pi}$ 
of the corresponding $\pi\pi\,\rightarrow\,\pi\pi$ partial wave,
\begin{eqnarray}
{\mathrm Im}\,{{F}^I_{J \Lambda_\gamma}}(\gamma\gamma\,\rightarrow\,\pi\pi)
&&\;=\;
{\rho_{\pi\pi}}\;
{{F}^{I \ast}_{J \Lambda_\gamma}}(\gamma\gamma\,\rightarrow\,\pi\pi{)}\;
{{\cal{I}}^I_{J}}(\pi\pi\,\rightarrow\,\pi\pi)\nonumber\\
&&\Downarrow\nonumber\\
{\phi^{I\;(\gamma\gamma\,\rightarrow\,\pi\pi)}_{J}}(t)&&\;=\;
{\delta^{IJ}_{\pi\pi}}(t)\,.
\end{eqnarray}
This fact can be incorporated into the so-called Omn\`{e}s function,
which is constructed to have the phase of the $\pi\pi$ scattering
amplitude above $\pi\pi$ threshold, and to be real otherwise,
\begin{equation}
{\Omega^I_J}(t)\;=\;\exp{\left[
\frac{t}{\pi}\;{\int_{4m_\pi^2}^{\infty}}\,dt'\,
\frac{{\delta^{IJ}_{\pi\pi}}(t')}{t'(t'\,-\,t\,-\,i\varepsilon)}\right]}
\;.
\end{equation}
The function ${{F}^I_{J \Lambda_\gamma}}{(\Omega^I_J{)^{-1}}}(t)$ is
by construction real above $\pi\pi$ threshold, 
but complex below threshold due to
the complexity of the Born partial waves ${B^{I}_{J\Lambda_\gamma}}$.
Hence we can write a dispersion relation for
$\left[{{F}^I_{J \Lambda_\gamma}}\;-\;{B^I_{J \Lambda_\gamma}}\right]
{(\Omega^I_J{)^{-1}}}(t) / t (t - 4 m_\pi^2)^{J/2}$,
 
\begin{eqnarray}
&&{{F}^I_{J \Lambda_\gamma}}(t)\;=\;\nonumber\\
&&{\Omega^I_J}(t)\;\left\{
{B^I_{J \Lambda_\gamma}}(t)\,{\mathrm Re}\left[(\Omega^I_J{)^{-1}}(t)\right]
\;-\;
\frac{t(t\,-\,4m^2_{\pi})^{J/2}}{\pi}\,{\int^{\infty}_{4m^2_{\pi}}}\,dt'\,
\frac{{B^I_{J \Lambda_\gamma}}(t')
\,{\mathrm Im}\left[{(\Omega^I_J{)^{-1}}}(t')\right]}
{t'(t'\,-\,4m^2_{\pi})^{J/2}(t'\,-\,t)}\right\} \;.
\label{eq:gagapipidisprel}
\end{eqnarray}
For $t > 4 m_\pi^2$, this integral is understood to be a principal value
integral, which we implement by subtracting the integrand at $t' = t$.
In this way we obtain a regular integral, which
can be performed without numerical problems, 

\begin{eqnarray}
&&{{F}^I_{J \Lambda_\gamma}}(t)\;=\;
{\Omega^I_J}(t)\;\left\{
{B^I_{J \Lambda_\gamma}}(t)\,\left(
{\mathrm Re}\!\left[(\Omega^I_J{)^{-1}}(t)\right]\;+\;
{\mathrm Im}\!\left[(\Omega^I_J{)^{-1}}(t)\right]\,
{1\over\pi}\,\ln(\frac{t}{4m^2_{\pi}}\,-1\,)
\right)\right.\nonumber\\
&&\left.\;-\;
\frac{t(t\,-\,4m^2_{\pi})^{J/2}}{\pi}\,{\int^{\infty}_{4m^2_{\pi}}}
\,\frac{dt'}{t'(t'-t)} \;
\left(\frac{{B^I_{J \Lambda_\gamma}}(t')
\,{\mathrm Im}\!\left[{(\Omega^I_J{)^{-1}}}(t')\right]}
{(t'\,-\,4m^2_{\pi})^{J/2}}\,-\,
\frac{{B^I_{J \Lambda_\gamma}}(t)
\,{\mathrm Im}\!\left[{(\Omega^I_J{)^{-1}}}(t)\right]}
{(t\,-\,4m^2_{\pi})^{J/2}} \right)
\right\} \;.
\end{eqnarray}
In our formalism, 
the s($J=0$)- and d($J=2$)-waves are unitarized. For the s- and d-wave
$\pi \pi$ phaseshifts, we use the solutions that were determined in
Ref.~\cite{Frog77}.
For the higher partial waves, the corresponding $\pi \pi$ phaseshifts
are rather small and are not known with good precision. 
Therefore, we will approximate all higher partial waves $(J \geq 4)$ 
by their Born contribution. 
The full amplitudes for the charged and neutral channels 
can then be cast into the forms

\begin{eqnarray}
&&{{F}^{(\pi^+ \pi^-)}_{\Lambda_\gamma}}(t,\theta_{\pi \pi})
\,=\,{{B}_{\Lambda_\gamma}}(t,\theta_{\pi \pi}) \nonumber\\
&&\hspace{1.2cm}+\;
\sum_{J=0,2}\!\sqrt{2J+1}\sqrt{\frac{(J-\Lambda_\gamma)!}
{(J+\Lambda_\gamma)!}}\left[
\sqrt{2\over 3}{{F}^{I=0}_{J \Lambda_\gamma}}(t)\,+\,
\sqrt{1\over 3}{{F}^{I=2}_{J \Lambda_\gamma}}(t)\,-\,
{{B}_{J \Lambda_\gamma}}(t)\right]
P_J^{\Lambda_\gamma} (\cos \theta_{\pi \pi}) \;,\\
&&{{F}^{(\pi^0 \pi^0)}_{\Lambda_\gamma}}(t,\theta_{\pi \pi})\,=\,
\sum_{J=0,2}\sqrt{2J+1}\sqrt{\frac{(J-\Lambda_\gamma)!}
{(J+\Lambda_\gamma)!}}\left[
-\sqrt{1\over 3}{{F}^{I=0}_{J \Lambda_\gamma}}(t)\,+\,
\sqrt{2\over 3}{{F}^{I=2}_{J \Lambda_\gamma}}(t)\right]
P_J^{\Lambda_\gamma} (\cos \theta_{\pi \pi}) \;.
\end{eqnarray}
\newline
\indent
The two-pion intermediate contribution holds to good precision up to 
$K \bar K$ threshold ($\approx$~1~GeV$^2$), because the four-pion
intermediate state couples only weakly and gives only small
inelasticities in the $\pi \pi$ phaseshifts. 
\newline
\indent
In Figs.~\ref{fig:gagapipitot} and \ref{fig:diffpippim}, we show our
results for the total and differential 
$\gamma\gamma\rightarrow\pi^+ \pi^-$ cross sections 
and a comparison to the existing data.  
In the threshold region, the charged pion cross sections are clearly
dominated by the Born graphs of Fig.~\ref{fig:born} because of the
vicinity of the pion pole in the $t$-channel of the 
$\gamma\gamma\rightarrow\pi^+ \pi^-$ process. 
However, the results for the 
unitarized calculation show that s-wave rescattering is not
negligible but leads to a considerable enhancement at energies just above
threshold. Besides the low energy structure, driven by the Born
terms, the $\gamma\gamma\rightarrow\pi \pi$ process has a prominent
resonance structure at higher energies corresponding to 
excitation of the isoscalar $f_2$(1270) resonance, with mass $m_{f_2}$
= 1275 MeV and width $\Gamma_{f_2}$ = 185.5 MeV \cite{PDG98}. 
The $f_2$ resonance shows up in the partial wave $F_{J=2 \,
  \Lambda_\gamma = 2}$ as outlined in Appendix~\ref{app:f2}. 
Therefore, the most efficient way to unitarize
this particular partial wave is to make a Breit-Wigner ansatz for 
the $f_2$ excitation, which is described in Appendix~\ref{app:f2} 
where we also give some details of the formalism for a spin-2
particle. The Breit-Wigner ansatz for the $f_2$ contribution to the
partial wave $F_{J=2 \, \Lambda_\gamma = 2}$ depends upon the
couplings $f_2 \pi \pi$ and $f_2 \gamma \gamma$. 
The coupling $f_2 \pi \pi$ is known from the decay 
$f_2 \rightarrow \pi \pi$ and is taken from Ref.~\cite{PDG98}. The 
coupling $f_2 \gamma \gamma$ is then fitted to the 
$\gamma\gamma\rightarrow\pi\pi$ cross section at the $f_2$ resonance
position, and is consistent with the value quoted in Ref.~\cite{PDG98}.  
The resulting amplitude, consisting of unitarized s-wave,
$f_2$ excitation and Born terms for all other partial waves (with $J
\geq 4$) is seen from Figs.~\ref{fig:gagapipitot} and
\ref{fig:diffpippim} to give a rather good description of the 
$\gamma\gamma\rightarrow\pi^+ \pi^-$ data up to $W_{\pi \pi} \simeq$
1.8 GeV. Only in the region $W_{\pi \pi} \approx$ 0.7 - 0.8 GeV, does
our description slightly overestimate the data.   
\newline
\indent
Having constructed the $\gamma\gamma\rightarrow\pi\pi$ amplitudes, we
next need the $\pi\pi \rightarrow N \bar N$ amplitudes 
in order to estimate the contribution of the $\pi \pi$ states to the
$t$-channel dispersion integral for Compton scattering. 
As we only kept s- and d-waves for $\gamma \gamma \rightarrow \pi
\pi$, we will only need the s- and d-waves ($J$ =
0, 2) for $\pi \pi \rightarrow N \bar N$. 
For every partial wave $J$, there are two independent 
$\pi \pi \rightarrow N \bar N$ helicity amplitudes 
$f_{\pm}^J(t)$, depending on 
whether the nucleon and anti-nucleon have the same ($f_{+}^J(t)$) or 
opposite ($f_{-}^J(t)$) helicities. We refer the reader to 
Appendix \ref{app:tchannel} (Eqs.~(\ref{eq:partialpipinnbar}) and 
(\ref{eq:frazerfulco})) for details. 
In this work, we take the s- and d-waves 
from the work of H\"ohler and collaborators~\cite{Hoehler83}, 
in which the lowest $\pi\pi \rightarrow N \bar N$ partial wave 
amplitudes were constructed from a partial wave solution of
pion-nucleon scattering, by use of the $\pi \pi$ phaseshifts of
Ref.~\cite{Frog77}, which we also used to construct the 
$\gamma \gamma \rightarrow \pi \pi$ amplitudes. 
In Ref.~\cite{Hoehler83}, the 
$\pi \pi \rightarrow N \bar N$ amplitudes are given 
for $t$ values up to $t \approx 40 \cdot m_\pi^2 \approx$
0.78~GeV$^2$, which will serve well for our purpose since the subtracted 
$t$-channel dispersion integral will have converged much below this
value as shown in the following. 
\newline
\indent
Finally, we can now combine 
the $\gamma \gamma \rightarrow \pi \pi$ and $\pi \pi \rightarrow N
\bar N$ amplitudes to construct the
discontinuities of the Compton amplitudes across the $t$-channel cut. 
In Appendix \ref{app:tchannel}, we show in detail how the Compton
invariant amplitudes $A_1,...,A_6$ are expressed by the
$t$-channel ($\gamma \gamma \rightarrow N \bar N$) helicity
amplitudes. Through unitarity we then express the imaginary parts of
these $t$-channel ($\gamma \gamma \rightarrow N \bar N$) helicity
amplitudes in terms of the $\gamma \gamma \rightarrow \pi \pi$ and 
$\pi \pi \rightarrow N \bar N$ amplitudes. We
finally express the discontinuities Im$_t A_i$ of the invariant
amplitudes $A_i$ ($i$ = 1,...,6) in terms of the corresponding 
$\gamma \gamma \rightarrow \pi \pi$ and $\pi \pi \rightarrow N
\bar N$ partial wave amplitudes (see Eq.~(\ref{eq:ima2pi})). 
As we restrict ourselves to s- and d-wave intermediate states 
in the actual calculations, we give here the expressions at
$\nu = 0$, including s- and d-waves only, 
that are needed for the subtracted $t$-channel dispersion
integral of Eq.~(\ref{eq:subt})~,

\begin{eqnarray}
{\mathrm Im}_t A_1 (\nu = 0, t)^{2 \pi} &=& - \; 
\sqrt{{t/4 - m_\pi^2} \over t} \; 
{1 \over {t \, (M^2 - t/4)}}\;
F_{0\,\Lambda_\gamma = 0}(t) \; f^{0*}_+(t) \; \nonumber\\
&& - \; \left( {{t/4 - m_\pi^2} \over t} \right)^{3/2} \; 
{\sqrt{5} \over 2}\;F_{2\,\Lambda_\gamma = 0}(t) \; f^{2*}_+(t) \;, \nonumber\\ 
{\mathrm Im}_t A_2 (\nu = 0, t)^{2 \pi} &=& 0 \;, \nonumber\\
{\mathrm Im}_t A_3 (\nu = 0, t)^{2 \pi} &=&  -  
\left( {{t/4 - m_\pi^2} \over t} \right)^{3/2} \, 
{{M^2} \over {(M^2 - t/4)}} \,
{\sqrt{5} \over 2} \; F_{2\,\Lambda_\gamma = 2}(t) \, 
\left\{ \sqrt{{3 \over 2}} \, f^{2*}_+(t) 
\,-\, M \, f^{2*}_-(t) \right\}\, , \nonumber\\
{\mathrm Im}_t A_4 (\nu = 0, t)^{2 \pi} &=& 0 \;, \nonumber\\
{\mathrm Im}_t A_5 (\nu = 0, t)^{2 \pi} &=&  -  
\left( {{t/4 - m_\pi^2} \over t} \right)^{3/2} \; M \;
\sqrt{{{15} \over 2}} \; 
F_{2\,\Lambda_\gamma = 0}(t) \; f^{2*}_-(t) \;, \nonumber\\
{\mathrm Im}_t A_6 (\nu = 0, t)^{2 \pi} &=&  -  
\left( {{t/4 - m_\pi^2} \over t} \right)^{3/2} \; M \;
{\sqrt{5} \over 2} \; F_{2\,\Lambda_\gamma = 2}(t) \; f^{2*}_-(t) \;. 
\label{eq:ima2pisd}
\end{eqnarray}
The reader should note that the s-wave $\pi \pi$ 
intermediate state only contributes
to the amplitude $A_1$. It is the $t$-dependence of 
this $I = J = 0$ $\pi \pi$ state in the
$t$-channel that is approximated 
in Ref.\cite{lvov97} and parametrized by a ``sigma'' pole.  
The d-wave $\pi \pi$ intermediate state gives rise to imaginary parts
for the amplitudes $A_1, A_3, A_5$ and $A_6$. The amplitude $A_2$ (at
$\nu = 0$) corresponds to the $t$-channel exchange of an object with 
the quantum
numbers of the pion (e.g. $\pi^0$ pole in Eq.~(\ref{eq:piopole})). 
Therefore two-pion intermediate states do not have the quantum numbers
to contribute to the amplitude $A_2$. 
The imaginary part of $A_4$ receives only contributions from
$\pi \pi$ intermediate states with $J \geq 4$ (see Eq.~(\ref{eq:ima2pi})) and
therefore is zero in our description, as we keep only s- and d-waves. 
\newline
\indent
In Fig.~\ref{fig:tchannelconv} we show the convergence of the 
$t$-channel integral from $4 m_\pi^2$ to $\infty$ 
in the subtracted DR of Eq.~(\ref{eq:subt}). We do so by calculating the
dispersion integral as function of the upper integration limit $t_{\rm
  upper}$ and by showing the ratio with the integral for 
$t_{\rm upper}$ = 0.78~GeV$^2$. The latter value corresponds to the highest
$t$ value for which the $\pi \pi \rightarrow N \bar N$ amplitudes are
given in Ref.~\cite{Hoehler83}. One clearly sees from 
Fig.~\ref{fig:tchannelconv} that the unsubtracted $t$-channel DR shows
only a slow convergence, whereas the subtracted $t$-channel DR has
already reached its final value, within the percent level, 
at a $t$ value as low as 0.4~GeV$^2$.

\section{Results and discussion}
\label{results}

In this section we shall present our results for Compton
scattering off the nucleon in the dispersion
formalism presented above.  

The real and imaginary parts of the six Compton amplitudes are
displayed in Fig.~\ref{fig:a1a6reim}. Note that for the real part, we
only show the subtracted $s$-channel integral of
Eq.~(\ref{eq:sub}). As can be seen from Fig.~\ref{fig:a1a6reim}, these
amplitudes show strong oscillations due to interference effects
between different pion photoproduction multipoles, in particular for
threshold pion production by $E_{0+}$ and $\Delta$-excitation by
$M_{1+}$. 

In Figs.~\ref{fig:threshold_gpi} and~\ref{fig:threshold_amb} we show our
predictions in the subtracted DR formalism and compare them with
the available Compton data on the proton 
below pion threshold. These data were used in
Ref.~\cite{McGibbon} to determine the scalar
polarizabilities $\alpha$ and $\beta$ through a global fit, 
with the results given in Eq.~(\ref{eq1.1}). In the analysis of
Ref.~\cite{McGibbon}, the unsubtracted DR formalism was used and the
asymptotic contributions (Eq.~(\ref{eq:aintas})) to the invariant
Compton amplitudes $A_1$ and $A_2$ were parametrized. In particular, 
$A^{as}_2$ was described by the $\pi^0$ pole, which yields the value
$\gamma_\pi \simeq - 45$. The free parameter entering in $A^{as}_1$
was related to $\alpha - \beta$, for which the
fit obtained the value $\alpha - \beta \simeq$ 10. 
Keeping $\alpha - \beta$ fixed at that value, we demonstrate in 
Fig.~\ref{fig:threshold_gpi} 
that the sensitivity to $\gamma_\pi$ is not at all negligible,
especially at the backward angles and the higher energies. 
Although we do not intend to give a best fit at 
the present stage, the subtracted DR formalism allows one 
to directly use the values $\alpha - \beta$ and
$\gamma_\pi$ as fit parameters, as is obvious from Eqs.~(\ref{eq:sub},
\ref{eq:subt}). We
investigate this further in Fig.~\ref{fig:threshold_amb}, where we show
our results for different $\alpha - \beta$ and 
for a fixed value of $\gamma_\pi = -37$, 
which is consistent with the heavy baryon ChPT
prediction \cite{HHKK} and close to the value obtained in Ref.~\cite{LN}
in a backward DR formalism. For that value of $\gamma_\pi$, 
a better description of the data (in particular at the backward angle)
seems to be possible by using a smaller value for $\alpha - \beta$
than determined in Ref.~\cite{McGibbon}. For a more reliable
extraction of the polarizabilities, 
more accurate data over the whole angular and energy
range are necessary. Recently, Compton data were taken at MAMI
over a wide angular range below pion threshold \cite{Olmos}. It will
be interesting to perform a best fit for $\alpha - \beta$ and
$\gamma_\pi$ with such an extended data base. 

As one moves to energies above pion threshold, the Compton cross section rises
rapidly because of the unitarity coupling to the much stronger
pion photoproduction channel. Therefore this higher energy region is 
usually considered less `pure' to extract polarizabilities because the
procedure would require a rather precise knowledge of pion
photoproduction. With the quite accurate pion photoproduction data 
on the proton that have become available in
recent years, the energy region above pion threshold for the Compton channel
could however serve as a valuable complement to determine 
the polarizabilities, provided 
one can minimize the model uncertainties in the dispersion formalism. 
In this work, we use the most recent information on the pion
photoproduction channel by taking the HDT \cite{HDT} 
multipoles at energies $E_\gamma \leq$ 500 MeV and 
the SAID-SP98K solution \cite{said} at higher energies. 
In addition, as previously shown in Fig.~\ref{fig:a1a2conv}, 
the subtracted dispersion relations are practically 
saturated by the one-pion channel for photon energies through 
the $\Delta$ region, which minimizes the uncertainty due to 
the modeling of the two-pion photoproduction channels. 
In Figs.~\ref{fig:compton_delta_gpi} and~\ref{fig:compton_delta_amb}
we display the sensitivity of the Compton cross sections to 
$\gamma_\pi$ and $\alpha - \beta$ in the lower part of the
$\Delta$ region in comparison with the available data. 
As can be seen from Fig.~\ref{fig:compton_delta_gpi}, these 
data are quite sensitive to the backward spin
polarizability $\gamma_\pi$. This sensitivity was exploited in
Ref.~\cite{Tonnison} within the context of an unsubtracted DR formalism, 
and the value $\gamma_\pi \simeq -27$ was extracted from the LEGS 97 
data, which are shown at the higher energies in 
Fig.~\ref{fig:compton_delta_gpi}. 
Our results for the subtracted DR are obtained in
Fig.~\ref{fig:compton_delta_gpi} by variation of $\gamma_\pi$ at
fixed $\alpha - \beta$ = 10, and by variation of $\alpha - \beta$ at
fixed $\gamma_\pi$ = -37 in Fig.~\ref{fig:compton_delta_amb}. 
For $\gamma_\pi$ we show the results for values between $\gamma_\pi = -27$
and $\gamma_\pi = -37$. One sees that the lower energy data
($E_\gamma$ = 149 MeV and 182 MeV) can be easily described by the larger
values of $\gamma_\pi$ if $\alpha - \beta$ decreases to some value below 10. 
On the other hand, the higher energy data ($E_\gamma$ = 230 MeV and 287
MeV) seem to favor a smaller value of $\gamma_\pi$, and so far we
confirm the conclusion reached in Ref.~\cite{Tonnison}. 
However, we have to point out that, at
these higher energies, the data around 90$^{\rm o}$ 
cannot be described in our subtracted DR formalism for reasonable
values of $\alpha - \beta$ and $\gamma_\pi$. 

This is also seen in Fig.~\ref{fig:fixed_th} at two fixed angles, now
as function of the energy throughout the $\Delta$ resonance region, 
for the MAMI data at 75$^{\rm o}$ \cite{Peise} and 90$^{\rm o}$ 
\cite{Molinari}. 
It is again obvious at these angles that the sensitivity to 
$\gamma_\pi$ is quite small. Therefore the physics of these data
is basically driven by pion photoproduction. With the multipoles
used here, the 75$^{\rm o}$ data are well described, but 
our prediction falls below the 90$^{\rm o}$ data 
on the left shoulder of the $\Delta$ resonance. 
In the same energy region there exist also both differential cross
section and photon asymmetry data from LEGS
\cite{Blanpied97} by use of the laser backscattering technique. 
In Fig.~\ref{fig:asymm} we compare our predictions with these data. 
One finds that at both energies ($E_\gamma$ = 265 MeV and 323 MeV) our
subtracted DR formalism provides a good description
of the asymmetries which display only little sensitivity on $\gamma_\pi$,
but underestimates the absolute values of the cross sections. In
particular close to the resonance position at $E_\gamma$ = 323 MeV,
the subtracted DR formalism does not allow us to find any reasonable
combination of $\gamma_\pi$ and $\alpha - \beta$ to describe these
data. Therefore, within the present subtracted DR formalism, 
the actual data situation at these higher energies 
does not seem to be conclusive to reliably decide on a value of
$\gamma_\pi$. Since the uncertainties due to two-pion and heavier
meson photoproduction are less than 1 \% in our subtracted DR
formalism, the only possibility to describe the $E_\gamma$ = 323 MeV
LEGS data would be an increase of the HDT $M_{1+}$ multipole by about
2.5 \% (see the dotted lines in Fig.~\ref{fig:asymm}). Indeed such a
fit was obtained by Tonnison et al. \cite{Tonnison} by use of the 
LEGS pion photoproduction multipole set of Ref.\cite{Blanpied97}
for photon energies between 200 and 350 MeV and the SAID-SM95 multipole
solution \cite{said} outside this interval. However the more recent 
SAID-SP98K solution is in very close agreement with the HDT multipoles
in the $\Delta$ region and hence the predictions also fall below the
data at 323 MeV in Fig.~\ref{fig:asymm}. 

Before coming to any conclusions, we like to point out that new Compton data
on the proton in and above the $\Delta$-resonance region 
and over a wide angular range 
have been measured recently at MAMI and reported preliminary \cite{LARA}. 
These new data will be most valuable to check if a systematic and
consistent trend becomes visible  between the data sets at the lower
energies and in the $\Delta$ region. 

Finally, in Fig.~\ref{fig:doublepol} we show that double polarization
observables will be ultimately necessary in order to reliably extract
the polarizabilities $\alpha - \beta$ and $\gamma_\pi$. 
In particular, an experiment with a circularly polarized photon 
and a polarized proton target displays quite some sensitivity on the 
backward spin polarizability $\gamma_\pi$, especially at the somewhat
higher energies of $E_\gamma \approx$~230 MeV. 
Such a measurement is more selective to $\gamma_\pi$ due to a lesser
sensitivity to $\alpha - \beta$ (see Fig.~\ref{fig:doublepol}). It
will be indeed a prerequisite to disentangle the scalar and vector
polarizabilities of the nucleon.

\section{Conclusions}
\label{conclusions}

In this work we have presented a formalism of fixed-$t$ subtracted
dispersion relations for Compton scattering off the nucleon at
energies $E_\gamma \leq$ 500 MeV. Due to the subtraction, the
$s$-channel dispersion integrals converge very fast and are 
practically saturated by the $\pi N$ intermediate states. 
Because of the use of subtracted DR we have minimized the
uncertainty from multi-pion and heavier meson intermediate states. 
Hence this formalism provides a direct cross-check between Compton
scattering and one-pion photoproduction. 
We have described this dominant 
$\gamma N \rightarrow \pi N \rightarrow \gamma N$
contribution by using the recent pion photoproduction multipoles of HDT. 
\newline
\indent
To calculate the functional dependence of the subtraction 
functions on the momentum transfer $t$, we
have included experimental information on the $t$-channel process 
through $\pi \pi$ intermediate states as 
$\gamma \gamma \rightarrow \pi \pi \rightarrow N \bar N$. We have constructed a
unitarized amplitude for the $\gamma \gamma \rightarrow \pi \pi$
subprocess and found a good description of available data. In this
way, we have largely avoided the uncertainties in Compton scattering 
associated with the two-pion continuum in the $t$-channel, which was
often modeled through the exchange of a somewhat fictitious $\sigma$-meson. 

As the polarizabilities directly 
enter as subtraction constants in the subtracted DR formalism, 
it can be used to extract the nucleon
polarizabilities from data with a minimum of model dependence. 
We have demonstrated the sensitivity to the polarizabilities 
$\alpha - \beta$ and $\gamma_\pi$ of existing Compton data on the 
proton both below pion threshold as well as in the $\Delta$ resonance region. 
The effects of the polarizabilities $\alpha - \beta$ and $\gamma_\pi$ 
on the Compton cross sections are strongly correlated.  
Hence these polarizabilities can only be determined
simultaneously from unpolarized observables even below pion
threshold. When comparing the
subtracted DR formalism with the existing data in the $\Delta$
resonance region, we have found that the actual data situation at
these higher energies does not seem to be conclusive to reliably
decide on a value for $\gamma_\pi$. However new Compton data on the
proton both below pion threshold and in the $\Delta$ region are
actually under analysis and will extend considerably the experimental data
base to fit the proton polarizabilities. 

We have also argued that double polarization observables
will ultimately be necessary to extract reliably $\alpha - \beta$ and 
$\gamma_\pi$. In particular we have shown that experiments of
polarized photons on polarized protons show rather large sensitivity
to $\gamma_\pi$ at energies around $E_\gamma \approx$ 230 MeV and at
backward angles. Therefore such polarization experiments hold the promise to
disentangle scalar and vector polarizabilities of the nucleon and 
to quantify the nucleon spin response in an external electromagnetic field.

\section*{acknowledgments}

The authors are grateful to G. Krein, A. L'vov 
and members of the A2-Collaboration, in particular J. Ahrens and 
V. Olmos de L\'eon, for useful discussions.
This work was supported by the Deutsche Forschungsgemeinschaft
(SFB 443) and in part by the Marie Curie TMR program (contract
ERBFMBICT972758).  

\newpage

\begin{appendix}

\section{The Mandelstam plane -- physical and  spectral regions for
 Compton scattering}
\label{app:spectralcompton}

The kinematics of Compton scattering, $\gamma(q)N(p)\rightarrow
\gamma(q')N(p')$, can be described in terms of the familiar Mandelstam
variables,

\eqa
\label{eqA1}
s=(q+p)^2\ ,\ \ t=(q-q')^2\ ,\ \ u=(q-p')^2\ ,
\eea
with the constraint
\eqa
\label{eqA2}
s+t+u=2M^2\ .
\eea
Furthermore we introduce the coordinate $\nu$ perpendicular to $t$,
\eqa
\label{eqA3}
\nu=\frac{s-u}{4M}=E_\gamma+\frac{t}{4M}\ .
\eea
In these equations, $E_\gamma$ is the photon energy in the $lab$ frame
and $M$ the nucleon mass.

The boundaries of the physical regions in the $s$, $u$ and $t$
channels are determined by the zeros of the Kibble function $\Phi$,

\eqa
\label{eqA4}
\Phi(s,t,u)=t(us-M^4)=0\ .
\eea
The 3 physical regions are shown by the horizontally hatched areas in
Fig.~1. The vertically hatched areas are the regions of non-vanishing
double spectral functions. These spectral regions are those regions
in the Mandelstam plane where two of the three variables 
$s, t$ and $u$ take on values that
correspond with a physical (i.e. on-shell) intermediate state. 
The boundaries of these regions follow from
unitarity. As discussed in Ref.~\cite{Hoehler83}, it is sufficient to
consider two-particle intermediate states in all channels. Since these
boundaries depend only on the masses, they are the same for all 6
amplitudes $A_i$. In the Mandelstam diagram of Fig.~1 they are
symmetric to the line $\nu=0$ due to crossing symmetry. For the
spectral function $\rho_{su}$ we obtain the boundary

\eqa
\label{eqA5}
b_I(u,s)=b_I(s,u)=[s-(M+m_\pi)^2]
[u-(M+m_\pi)^2]
-(m_\pi^2 + 2 \,M \,m_\pi)^2=0\ ,
\eea
and for the spectral function $\rho_{st}$ we find

\eqa
\label{eqA6}
b_{II}(s,t)=
(t-4 m_\pi^2)[s-(M+m_\pi)^2]
[s-(M-m_\pi)^2]-8m_\pi^4(s+M^2-m_\pi^2/2)=0
\eea
The boundary of the spectral function $\rho_{ut}$ follows from
crossing symmetry. We also note that these boundaries are obtained for
the isovector photon, which couples to a $\pi^+\pi^-$ pair. The
corresponding boundaries for the isoscalar photon are inside the
boundaries of Eqs.~(\ref{eqA5}) and (\ref{eqA6}), because it couples to
3 pions.

\section{{\it t}-channel helicity amplitudes for Compton scattering}
\label{app:tchannel}

The $t$-channel helicity amplitudes for Compton scattering 
can be expressed in the orthogonal basis of Prange \cite{Prange58} 
in the following form:

\begin{eqnarray}
T_{\lambda_N \lambda_{\bar{N}},\,\lambda'_{\gamma}\lambda_{\gamma}}^t\,
(\nu, t)\;&=&\;(-1)^{\frac{1}{2}-\lambda_{\bar{N}}}\;
{\varepsilon'}_{\mu}(q',{\lambda'}_{\gamma}) \;
{\varepsilon}_{\nu}(q,\lambda_{\gamma}) \nonumber\\
&&\times \;{\bar u}(\vec p~', \lambda_N)\, \left\{ 
\,- \, {{\tilde{P'}^{\mu}\,\tilde{P'}^{\nu}} \over {\tilde{P'}^{2}}} \; 
\left(\, T_{1} \,+\, \tilde{K}\!\!\!\!/ \,T_{2}\, \right) \right.\nonumber\\ 
&&\hspace{2.5cm}- \, {{\tilde{N}^{\mu}\,\tilde{N}^{\nu}}\over \tilde{N}^{2}}\,
\left(\, T_{3} \,+\, \tilde{K}\!\!\!\!/ \,T_{4}\,\right) \nonumber\\
&&\hspace{2.5cm} +\,i\,
{{\tilde{P'}^{\mu}\tilde{N}^{\nu}\,-\,\tilde{P'}^{\nu}\tilde{N}^{\mu}}\over
{\tilde{P'}^{2}\,\tilde{K}^{2}}} \,\gamma_{5}\,T_{5} \nonumber\\
&&\hspace{2.5cm} \left. +\,i\, 
{{\tilde{P'}^{\mu} \tilde{N}^{\nu}\,+\,\tilde{P'}^{\nu} \tilde{N}^{\mu}} \over
{\tilde{P'}^{2}\,\tilde{K}^{2}} }\,\gamma_{5}\,\tilde{K}\!\!\!\!/ \,T_{6} 
\right\}\;v(- \vec p, \lambda_{\bar N}) \; ,
\label{eq:thelcov}
\end{eqnarray}
where
\begin{eqnarray}
\tilde{P'}^{\mu} &=& \tilde{P}^{\mu}\,-\,\tilde{K}^{\mu}\,
{{\tilde{P} \cdot \tilde{K}} \over {\tilde{K}^2}} \;, 
\hspace{0.5cm} \tilde{P} = {1 \over 2}\,(-\,p \,+\, p')\;, 
\hspace{0.5cm} \tilde{K} = {1 \over 2}\,(q \,-\, q'),\nonumber\\
\tilde{N}^{\mu}&=&{\epsilon}^{\mu \alpha \beta \gamma}\,
\tilde{P'}_{\alpha}\,\tilde{Q}_{\beta}\,\tilde{K}_{\gamma} \;, \hspace{.5cm} 
\tilde{Q} = {1 \over 2}\,(-\,p \,-\, p')\,=\,{1 \over 2}\,(-\,q' \,-\, q)\, ,
\end{eqnarray}
and using the convention $\epsilon_{0 1 2 3} = + 1$.
\newline
\indent
In the c.m. system of the $t$-channel process $\gamma \gamma
\rightarrow N \bar N$ (see Fig.~\ref{fig:tkinematics} for the
kinematics), we choose the photon momentum $\vec q_t$
(helicity $\lambda'_{\gamma}$) to point in the $z$-direction and the
nucleon momentum $\vec p~' = \vec p_t$ in the $xz$ plane at an angle
$\theta_t$ with respect to the $z$-axis (the anti-nucleon momentum is
then given by $- \vec p = - \vec p_t$).  In this frame, the helicity
amplitudes of Eq.(\ref{eq:thelcov}) can be cast into the form
\begin{eqnarray}
T_{\lambda_N \lambda_{\bar{N}},\,\lambda'_{\gamma}\lambda_{\gamma}}^t\,
(\nu, t) 
\;=\;(-1)^{\frac{1}{2}-\lambda_{\bar{N}}}\,\bar{u}(\vec{p_t},\lambda_N)\;
&&\left\{ \,-\,\frac{1}{2}\,{\lambda'}_{\gamma} \lambda_{\gamma}
\,\left(T_1 \,+\,|\vec{q_t}|\,\gamma^3 \, T_2\right) \right.\nonumber\\
&&\hspace{.4cm}-\;{1 \over 2}\;
\left( T_3 \,+\,|\vec{q_t}|\,\gamma^3 \,T_4\right) \nonumber\\
&&\hspace{0.4cm}-\,
\frac{1}{2}\;({\lambda'}_{\gamma} + \lambda_{\gamma}) 
\;\gamma_5 \,T_5 \nonumber\\
&&\left.\hspace{.4cm}-\,\frac{1}{2}\;({\lambda'}_{\gamma} - \lambda_{\gamma})
\,\gamma_5 \,|\vec{q_t}|\,\gamma^3\,T_6 \;\right\}\;
v(-\vec{p_t},\lambda_{\bar{N}}).
\end{eqnarray}
Under parity transformation, these amplitudes behave as
\begin{eqnarray}
T_{\lambda_N\lambda_{\bar{N}},\,\lambda'_{\gamma}\lambda_{\gamma}}^t\,(\nu, t) 
\;=\; (-1)^{\Lambda_{N} - \Lambda_{\gamma}}\;
T_{-\lambda_N-\lambda_{\bar{N}},\,-\lambda'_{\gamma}-\lambda_{\gamma}}^t\,
(\nu, t) \;,
\end{eqnarray}
with the helicity differences $\Lambda_\gamma$ and $\Lambda_N$ given
by $\Lambda_\gamma = \lambda'_\gamma - \lambda_\gamma$ 
(with $\Lambda_\gamma$ = 0 or 2) and $\Lambda_N
= \lambda_N - \lambda_{\bar N}$ (with $\Lambda_N$ = 0 or 1) respectively. 

However, the invariant amplitudes $T_i (i = 1,...,6)$ of Prange have
the disadvantage to behave differently under $s \leftrightarrow u$
crossing. While $T_1$, $T_3$, $T_5$ and $T_6$ are even functions of
$\nu$, $T_2$ and $T_4$ are odd functions (note that $\nu
\rightarrow -\nu$ is equivalent to $s \leftrightarrow u$). Therefore,
L'vov used a new set of invariant amplitudes $A_i (i =
1,...,6)$, which are all even functions of $\nu$ and at the same 
time free of kinematical singularities~\cite{lvov97}
\begin{eqnarray}
A_1 &=& {1 \over t}\; \left[ T_1\;+\;T_3\;+\;\nu\,(T_2\;+\;T_4)\right]
\;, \nonumber\\
A_2 &=& {1 \over t}\; \left[ 2\,T_5\;+\;\nu\,(T_2\;+\;T_4)\right] \;,
\nonumber\\
A_3 &=& {M^2 \over {M^4 - su}} \;\left[ T_1\;-\;T_3\;-\;
{t \over {4 \nu}}\,(T_2\;-\;T_4)\right] \;,
\nonumber\\
A_4 &=& {M^2 \over {M^4 - su}} \;\left[ 2\,M\,T_6\;-\;
{t \over {4 \nu}}\,(T_2\;-\;T_4)\right] \;,
\nonumber\\
A_5 &=& {1 \over {4\nu}}\; \left[ T_2\;+\;T_4 \right] \;,\nonumber\\
A_6 &=& {1 \over {4\nu}}\; \left[ T_2\;-\;T_4 \right] \;.
\label{eq:aifuncti}
\end{eqnarray}
In terms of the $t$-channel helicity amplitudes $T_{\lambda_N
  \lambda_{\bar N}, \, \lambda'_{\gamma} \lambda_\gamma}^t(\nu, t)$ of
Eq.~(\ref{eq:thelcov}) the invariant amplitudes $A_i (\nu, t)$ ($i$ =
1,...,6) of Eq.~(\ref{eq:aifuncti}) read
\begin{eqnarray}
A_1 &=& {{1} \over{t \, \sqrt{t - 4 M^2}}} 
\left\{ \; \left[ T_{{1 \over 2}\,{1 \over 2},1\,1}^t \;+\;
T_{{1 \over 2}\,{1 \over 2},-1\,-1}^t \right]
\;-\; {{2 \, \nu \, \sqrt{t}} \over {\sqrt{su - M^4}}}\; 
T_{{1 \over 2}\,-{1 \over 2},1\,1}^t \right\} \;, \nonumber \\
A_2 &=& {{1} \over{t \, \sqrt{t}}} 
\left\{ \;- \left[ T_{{1 \over 2}\,{1 \over 2},1\,1}^t \;-\;
T_{{1 \over 2}\,{1 \over 2},-1\,-1}^t \right]
\;-\; {{2 \, \nu \, \sqrt{t - 4 M^2}} \over {\sqrt{su - M^4}}}\; 
T_{{1 \over 2}\,-{1 \over 2},1\,1}^t \right\} \;, \nonumber \\
A_3 &=& {{M^2} \over {su - M^4}} \; {{1} \over{\sqrt{t - 4 M^2}}} 
\left\{ \;2 \; T_{{1 \over 2}\,{1 \over 2},1\,-1}^t 
\;+\; {{\sqrt{su - M^4}} \over {\nu \,\sqrt{t}}} 
\;\left[ T_{{1 \over 2}\,-{1 \over 2},1\,-1}^t \;+\;
T_{{1 \over 2}\,-{1 \over 2},-1\,1}^t \right] \right\} \;, \nonumber\\
A_4 &=& {{M^2} \over {su - M^4}} \; {{1} \over{\sqrt{su - M^4}}} 
\left\{ \;M \left[- \; T_{{1 \over 2}\,-{1 \over 2},1\,-1}^t 
\;+\;T_{{1 \over 2}\,-{1 \over 2},-1\,1}^t \right] \right. \nonumber\\
&&\hspace{3.5cm}\left.
\;+\; {{\sqrt{t} \, \sqrt{t - 4 M^2}} \over {4 \, \nu}} 
\;\left[ T_{{1 \over 2}\,-{1 \over 2},1\,-1}^t \;+\;
T_{{1 \over 2}\,-{1 \over 2},-1\,1}^t \right] \right\} \;, \nonumber\\
A_5 &=& {{\sqrt{t - 4 M^2}} \over {4 \, \nu \, \sqrt{t} \, \sqrt{su - M^4}}} 
\left\{ \;-2 \; T_{{1 \over 2}\,-{1 \over 2},1\,1}^t \right\} \;, \nonumber\\
A_6 &=& {{\sqrt{t - 4 M^2}} \over {4 \, \nu \, \sqrt{t} \, \sqrt{su - M^4}}} 
\left\{ \; \left[ T_{{1 \over 2}\,-{1 \over 2},1\,-1}^t \;+\;
T_{{1 \over 2}\,-{1 \over 2},-1\,1}^t \right] \right\} \;.
\label{eq:thelampl}
\end{eqnarray}

In the subtracted DR of Eq.~(\ref{eq:subt}), the
$t$-channel integral runs along the line $\nu$ = 0. Therefore, we have
to determine the imaginary parts Im$_t A_i (\nu = 0, t)$ of the
invariant amplitudes of Eq.~(\ref{eq:thelampl}).  We start by
decomposing of the $t$-channel helicity amplitudes for $\gamma \gamma
\rightarrow N \bar N$ into a partial wave series,
\begin{equation}
T_{\lambda_N \lambda_{\bar N}, \, \lambda'_{\gamma} \lambda_\gamma}^t 
(\nu, t) \;=\; \sum_J {{2 J + 1} \over {2}} \;
T_{\lambda_N \lambda_{\bar N}, \, \lambda'_{\gamma}
  \lambda_\gamma}^{J}
(t)\; d^J_{\Lambda_N \Lambda_\gamma} (\theta_t) \;,
\label{eq:pwgagannbar}
\end{equation}  
where $d^J_{\Lambda_N \Lambda_\gamma}$ are Wigner $d$-functions and
$\theta_t$ is the scattering angle in the $t$-channel, which is
related to the invariants $\nu$ and $t$ by
\begin{equation}
\cos \theta_t = {{4 \, M \, \nu} \over {\sqrt{t} \, \sqrt{t - 4 M^2}}}\;. 
\label{eq:costht}
\end{equation}
It is obvious from this equation that $\nu = 0$ 
corresponds to $90^{\rm o}$ scattering for the $t$-channel process. 
As explained in Section \ref{tchannel}, we calculate 
the imaginary parts of the $t$-channel helicity amplitudes 
$T_{\lambda_N \lambda_{\bar N}, \, \lambda'_{\gamma} \lambda_\gamma}^t
(\nu, t)$ through the unitarity equation by inserting $\pi \pi$
intermediate states, which should give the dominant
contribution below $K \bar K$ threshold,
\begin{equation}
2\,{\mathrm Im} T_{\gamma \gamma \, \rightarrow \, N \bar N} \;=\; 
{1 \over { {(4 \pi)}^2 } } 
\;{ | {\vec {p}}_{{\pi}} | \over {\sqrt {t}} } 
\; \int {d \Omega}_{{\pi}}\,
\left [ \, T_{\gamma \gamma \, \rightarrow \, \pi \pi}\, \right ] 
\; \cdot \; 
\left [ \, T_{\pi \pi \, \rightarrow \, N \bar N}\, \right ] ^ { \ast }.
\end{equation}
Combining the partial wave expansion for $\gamma \gamma \rightarrow \pi
\pi$~,
\begin{equation}
T^{\gamma \gamma \, \rightarrow \, \pi \pi}
_{{\Lambda}_{\gamma}} (t, \theta_{\pi \pi}) \;=\;
\sum_{J even} \, {{2J+1} \over 2} \; 
T^{J \;(\gamma \gamma \, \rightarrow \, \pi \pi)}
_{{\Lambda}_{\gamma}}(t) \, \cdot \,
\sqrt {(J-\Lambda_{\gamma})! \over {(J+\Lambda_{\gamma})!}} \, \cdot \,
P_J^{\Lambda_{\gamma}}(\cos \theta_{\pi \pi}),
\label{eq:partialgagapipi}
\end{equation}
and the partial wave expansion for $\pi \pi \rightarrow N \bar N$,
\begin{equation}
T^{\pi \pi \, \rightarrow \, N \bar N}
_{\Lambda_N} (t, \Theta) \;=\;
\sum_{J} \, {{2J+1} \over 2} \; 
T^{J \; (\pi \pi \, \rightarrow \, N \bar N)}
_{\Lambda_N}(t) \, \cdot \,
\sqrt {(J-\Lambda_N)! \over {(J+\Lambda_N)!}} \, \cdot \,
P_J^{\Lambda_N}(\cos \Theta) \;.
\label{eq:partialpipinnbar}
\end{equation}
We can now construct the imaginary parts of the
Compton $t$-channel partial waves, 
\begin{equation}
2\,{\mathrm Im} T^{J \;(\gamma \gamma \, \rightarrow \, N \bar N)}
_{\lambda_N \lambda_{\bar N}, \, \lambda'_{\gamma} \lambda_\gamma}(t)\;=\; 
{1 \over {(8 \pi)} } 
\; {p_{\pi} \over \sqrt{t}}
\left [ \, T^{J \;(\gamma \gamma \, \rightarrow \, \pi \pi)}
_{\Lambda_\gamma}(t)\, \right ] 
\left [ \, T^{J \;(\pi \pi \, \rightarrow \, N \bar N)}
_{\Lambda_N }(t)\, \right ] ^ { \ast }.
\label{eq:partialtunit}
\end{equation}
The partial wave amplitudes $T^{J \;(\pi \pi \, \rightarrow \, N
  \bar N)} _{\Lambda_N}$ of
Eq.~(\ref{eq:partialpipinnbar}) are related to the
amplitudes $ f^J_{\pm}(t)$ of Frazer and Fulco
\cite{Frazer60} by the relations
\begin{eqnarray}
T^{J \;(\pi \pi \, \rightarrow \, N \bar N)}_{\Lambda_N = 0}(t)\;
&&= \frac{16 \pi}{p_N}\,(p_N \; p_\pi)^J \, \cdot \,
f^J_+ (t)\ ,\nonumber\\
T^{J \;(\pi \pi \, \rightarrow \, N \bar N)}_{\Lambda_N = 1}(t)\;
&&= 8 \pi \,\frac{\sqrt {t}}{p_N}\,(p_N \; p_\pi)^J \, \cdot \,
f^J_- (t)\ ,
\label{eq:frazerfulco}
\end{eqnarray}
with $p_N$ and $p_{\pi}$ the c.m. momenta of nucleon and pion
respectively,
\begin{equation}
p_N = \sqrt{t/4 - M^2} \;, \hspace{2cm} p_\pi = \sqrt{t/4 - m_\pi^2} \;.
\end{equation}
For the reaction $\gamma \gamma \, \rightarrow \, \pi \pi$, we will
use the partial wave amplitudes 
$F_{J\,\Lambda_\gamma}(t)$, which are
related to those of Eq.~(\ref{eq:partialgagapipi}) by
\begin{equation}
T^{J \;(\gamma \gamma \, \rightarrow \, \pi \pi)} _{\Lambda_\gamma}(t)
\;=\; \frac{2}{\sqrt{2J+1}} \cdot F_{J\,\Lambda_\gamma}(t) \;.
\label{eq:deffgagapipi}
\end{equation}
Denoting the Born partial wave amplitudes for
$\gamma \gamma \;\rightarrow \; {\pi}^+ {\pi}^-$ by 
$B_{J \, \Lambda_\gamma}(t)$, 
the lowest Born partial waves ($s$ and $d$ waves) are
\begin{eqnarray}
{B_{00}}(t)&&\;=\;2e^2\;{{1-{\beta}^2}\over {2 \beta}}\;
\ln{\left({{1+\beta}\over {1-\beta}}\right)}, \nonumber\\
{B_{20}}(t)&&\;=\;2e^2\;{\sqrt{5}\over 4}\;
{{1-{\beta}^2}\over{\beta^2}}\;\left\{ \,{{3-{\beta}^2}\over \beta}\,
\ln{\left({{1+\beta}\over {{1-\beta}}}\right)} \;-\;6\; \right\}, \nonumber\\
{B_{22}}(t)&&\;=\;2e^2\;{\sqrt{15}\over {4\sqrt{2}}}\;
\left\{ \,{(1-{\beta}^2)^2 \over {\beta^3}}\,
\ln{\left({{1+\beta}\over {1-\beta}}\right)} \;+\;{10 \over 3}\;
-\;{2\over {\beta}^2}\; \right\} \;,
\label{eq:gagapipiborn}
\end{eqnarray}
with $\beta$ = $ p_{\pi} / \left(\sqrt{t}/2 \right)$ the pion
velocity. 

Inserting the partial-wave expansion of Eq.~(\ref{eq:pwgagannbar}) into
Eq.~(\ref{eq:thelampl}), we can finally express the $2 \pi$ $t$-channel
contributions Im$_t A_i (\nu = 0, t)^{2 \pi}$ by the partial wave
amplitudes for the reactions $\gamma \gamma \, \rightarrow \, \pi \pi$
and $\pi \pi \, \rightarrow \, N \bar N$,
\begin{eqnarray}
&&{\mathrm Im}_t A_1 (\nu = 0, t)^{2 \pi} = {{p_\pi} \over {\sqrt{t}}} \, 
{1 \over {t \, p_N^2}}\,
\sum_{J = 0,2,4,...} \,(p_\pi \,p_N)^J \,\sqrt{2 J + 1} \,
F_{J\,\Lambda_\gamma = 0}(t) \, f^{J*}_+(t) \,
\left[ (-1)^{J/2} \, {{(J - 1)!!} \over {J!!}}\right] \, , \nonumber\\
&&{\mathrm Im}_t A_2 (\nu = 0, t)^{2 \pi} = 0 \;, \nonumber\\
&&{\mathrm Im}_t A_3 (\nu = 0, t)^{2 \pi} = \nonumber\\ 
&& {{p_\pi} \over {\sqrt{t}}} \; 
{{M^2} \over {t \, p_N^4}}\,
\sum_{J = 2,4,...} \,(p_\pi \,p_N)^J \;\sqrt{2 J + 1} \;\;
F_{J\,\Lambda_\gamma = 2}(t) \; 
\left[\, {(-1)^{(J - 2)/2}\;{\sqrt{(J + 1)J \over {(J - 1)(J + 2)}}}}
\; {{(J - 1)!!} \over {J!!}} \right] \nonumber\\
&&\hspace{3.5cm} \times \left\{ \;f^{J*}_+(t) 
\;-\; \;f^{J*}_-(t) \; {M}
\left[{ {(J + 2)(J - 1) \,-\,2} \over {\sqrt{J(J + 1)}}} 
\right] \right\} \;,\nonumber\\
&&{\mathrm Im}_t A_4 (\nu = 0, t)^{2 \pi} = \nonumber\\
&&- \,{{p_\pi} \over {\sqrt{t}}} \; {{M^3} \over {t \, p_N^4}}\,
\sum_{J = 4,...} \,(p_\pi \,p_N)^J \,\sqrt{2 J + 1} \,
F_{J\,\Lambda_\gamma = 2}(t)  \, f^{J*}_-(t) \,
{ {(-1)^{(J - 2)/2} \;2\; (J - 2)(J + 3)} \over \sqrt{(J + 2)(J - 1)}}
 \, {{(J - 1)!!} \over {J!!}} \, ,\nonumber\\
&&{\mathrm Im}_t A_5 (\nu = 0, t)^{2 \pi} = \nonumber\\ 
&&-\,{{p_\pi} \over {\sqrt{t}}} \; {M \over {t \, p_N^2}}\;
\sum_{J = 2,4,...} \,(p_\pi \,p_N)^J \;\sqrt{2 J + 1} \;\;
F_{J\,\Lambda_\gamma = 0}(t) \; f^{J*}_-(t) \; 
\left[ {{(-1)^{(J - 2)/2}} \over \sqrt{J(J + 1)} } \; 
{{(J + 1)!!} \over {(J - 2)!!}}\right] \;, \nonumber\\
&&{\mathrm Im}_t A_6 (\nu = 0, t)^{2 \pi} = \nonumber\\
&&-\,{{p_\pi} \over {\sqrt{t}}} \; {M \over {t \, p_N^2}}\;
\sum_{J = 2,4,...} \,(p_\pi \,p_N)^J \;\sqrt{2 J + 1} \;\;
F_{J\,\Lambda_\gamma = 2}(t) \; f^{J*}_-(t) \;\nonumber\\
&&\hspace{3.5cm} \times 
\left[\, { {(-1)^{(J - 2)/2}  \;\; \left[(J + 2)(J - 1) \,-\, 2\right]} 
\over  {\sqrt{(J + 2)(J - 1)}}}
 \; {{(J - 1)!!} \over {J!!}} \right] \;.
\label{eq:ima2pi}
\end{eqnarray}
We note that the s-wave $(J=0)$ component of the $2\pi$ intermediate
states contributes only to $A_1$.  The amplitude $A_2$,
corresponding to the exchange of pseudoscalar mesons (dominantly
$\pi^0$) in the $t$-channel, gets no contribution from $2\pi$ states,
because the $2 \pi$ system cannot couple to the nucleon through a
pseudoscalar operator. Furthermore, it is found that only waves with
$J\ge4$ contribute to the amplitude $A_4$. In our calculations we
saturate the $t$-channel dispersion integral with s(J = 0)- and 
d($J = 2$)-waves, for which the expressions of Eq.~(\ref{eq:ima2pi})
reduce to those given in Eq.~(\ref{eq:ima2pisd}).

\section{${\it F}_2\,$-meson contribution to the 
$\gamma\gamma\,\rightarrow\,\pi\pi$ process}
\label{app:f2}

A particle with mass $m$ and spin-2 is described in terms of a
symmetric and traceless field tensor $\Phi^{\mu\nu}$ satisfying the
Klein-Gordon equation.  Furthermore, as for a massive spin-1 field,
the `Lorentz gauge' condition requires that the four-divergence with
respect to one of the four-vector indices vanishes,
\begin{eqnarray}
\left(\,\Box \;+\; {m^2}\, \right)\,{\Phi^{\mu\nu}}&&\;=\;0 \;,\nonumber\\
{\Phi^{\mu\nu}}&&\;=\;{\Phi^{\nu\mu}} \;,\nonumber\\
{{\Phi^{\mu}}_{\mu}}&&\;=\;0 \;, \nonumber\\
{\partial_{\mu}}\,{\Phi^{\mu\nu}}&&\;=\;
{\partial_{\nu}}\,{\Phi^{\mu\nu}}\;=\;0\;.
\end{eqnarray}
Therefore, the tensor $\Phi^{\mu \nu}$ has only five independent
components.

A state of spin-2 is characterized by its polarization tensor
${\varepsilon^{\mu\nu}}\,(p, \Lambda),\;{\mathrm where}\; \Lambda
\;=\;\{2,1,0,-1,-2\}$ defines the polarization,
\begin{eqnarray}
{\varepsilon^{\mu\nu}}\,(p, \Lambda)&&\;=\;
{\varepsilon^{\nu\mu}}\,(p, \Lambda) \;,\nonumber\\
{{\varepsilon^{\mu}}_{\mu}}\,(p, \Lambda)&&\;=\;0 \;,\nonumber\\
{p_{\mu}}\,{\varepsilon^{\mu\nu}}\,(p, \Lambda)&&\;=\;
{p_{\nu}}\,{\varepsilon^{\mu\nu}}\,(p, \Lambda)\;=\;0\;.
\end{eqnarray}
The polarization tensor 
${\varepsilon^{\mu\nu}}\,(p, \Lambda)$ can be constructed from
(massive) spin-1 polarization vectors by
\begin{equation}
{\varepsilon^{\mu\nu}}\,(p, \Lambda)\;=
\;{\sum_{\lambda=-1,0,1}}\;{\sum_{\lambda'=-1,0,1}}\;
\langle\;1\,\lambda,\;1\,\lambda'\;|\;2\,\Lambda\;\rangle\; 
{\epsilon^{\mu}}\,(p,\lambda)\;{\epsilon^{\nu}}\,(p,\lambda') \;.
\label{eq:spin2pol}
\end{equation}
If one chooses the $z$-axis along the momentum of the particle, the three
polarization vectors of a massive spin-1 particle are
\begin{eqnarray}
&&{\epsilon^{\mu}}\,(p,\lambda\,=\,+1)\;=\;
(\;0\,,\,-\frac{1}{\sqrt{2}}\,,\,-\frac{i}{\sqrt{2}}\,,\,0\;) \;,\nonumber\\
&&{\epsilon^{\mu}}\,(p,\lambda\,=\,-1)\;=\;
(\;0\,,\,+\frac{1}{\sqrt{2}}\,,\,-\frac{i}{\sqrt{2}}\,,\,0\;) \;,\nonumber\\
&&{\epsilon^{\mu}}\,(p,\lambda\,=\,0)\;=\;
(\;|\vec{p}|,\;0,\;0,\;{p^0}\;)\;/\;m \;, 
\end{eqnarray}
where $p^{\mu}$ = $(p^0, 0, 0, |\vec{p}|)$ and $p^2 = m^2$. 
In the case of the spin-2 polarization tensor constructed 
as in Eq.~(\ref{eq:spin2pol}), the following normalization 
and completeness conditions hold :
\begin{eqnarray}
{\varepsilon^{\mu\nu}}\,(p, \Lambda)\;\cdot\;
{\varepsilon^{\ast}_{\mu\nu}}\,(p, \Lambda')\;&=&\;
\delta_{\Lambda\Lambda'} \;,\nonumber\\
{\sum_{\Lambda}}\;{\varepsilon^{\mu\nu}}\,(p, \Lambda)\;
{\big[\,{\varepsilon^{\alpha\beta}}\,(p, \Lambda)\,\big]^{\ast}}
\;&=&\;\frac{1}{2}\;\left(\;
{K^{\mu\alpha}}\,{K^{\nu\beta}}\;+\;
{K^{\nu\alpha}}\,{K^{\mu\beta}}\;-\;
\frac{2}{3}\,{K^{\mu\nu}}\,{K^{\alpha\beta}}\;\right) \;,\nonumber\\
{\mathrm with}\;\;{K^{\mu\nu}}\;=\;
-\,{g^{\mu\nu}}\;+\;\frac{{p^{\mu}}{p^{\nu}}}{m^2} \;.
\end{eqnarray}
Finally, the propagator of the spin-2 particle with total width
$\Gamma$ takes the form
\begin{equation}
i \;{\Delta^{\mu\nu\alpha\beta}}\,(p)\;=\;
{{{ i \; \sum_{\Lambda}}\;\big[\,{\varepsilon^{\mu\nu}}\,(p, \Lambda)
\,\big]^{\ast}\;{{\varepsilon^{\alpha\beta}}\,(p, \Lambda)}}\over
{{p^2}\;-\;{m^2}\;+\;i\,m\,{\Gamma}}}\ .
\end{equation}

We will now apply this spin-2 formalism to describe the $s$-channel
exchange of the $f_2$ meson in the process $\gamma \gamma \rightarrow
\pi \pi$.
\newline
\indent 
The coupling of the (isoscalar) $f_2$(1270) meson to a 
pion pair (with momenta $p_\pi$, $p'_\pi$ and cartesian isospin
indices $a, b$) is described by the amplitude 
\begin{equation}
{\cal{M}}\left({f_2}\,\rightarrow\,\pi \pi\right)
\;=\;\frac{g_{{f_2}\pi\pi}}{m_{f_2}} \; \delta_{a b}
\;{p'_\pi}^{\mu}\;p_\pi^{\nu}\;{{\varepsilon}_{\mu\nu}}\,(p, \Lambda)\;,
\end{equation}
where $p$ is the $f_2$ meson momentum and $m_{f_2}$ its mass. 
The coupling constant $g_{{f_2}\pi\pi}$ is determined from the
${f_2}\,\rightarrow\,\pi \pi$ decay width~:
\begin{equation}
\Gamma({f_2}\,\rightarrow\,\pi \pi)\;=\;{1 \over 40 \pi}\;{g_{{f_2}\pi\pi}^2}\;
\frac{\left(p_\pi\right)^5}{m_{f_2}^4}\ ,
\label{eq:f2pipiw}
\end{equation}
where $p_\pi = \sqrt{m_{f_2}^2 / 4 - m_\pi^2}$ 
is the pion momentum in the $f_2$ rest frame.
Using the partial width 
$\Gamma({f_2}\,\rightarrow\,\pi \pi)\;=\;0.846\,\Gamma_0$ and
the total $f_2$-width $\Gamma_0\;=\;185$~MeV \cite{PDG98},
Eq.~(\ref{eq:f2pipiw}) yields for the coupling : 
$g_{{f_2}\pi\pi} \simeq$~23.64.
\newline
\indent
The Lorentz structure of the vertex 
${f_2}\,\rightarrow\,\gamma \gamma$ is given by
\begin{equation}
{\cal{M}}\left({f_2}\,\rightarrow\,\gamma \gamma\right)\;=\;-i \;2\,
e^2 \, \frac{g_{{f_2}\gamma \gamma}}{m_{f_2}}
\;{{\cal{F}}^{\mu\delta}}\left(q, \lambda_\gamma \right)
{{{\cal{F}}_{\delta}}^{\nu}}\left(q', \lambda'_\gamma\right)\;
{{\varepsilon}_{\mu\nu}}\,(p, \Lambda)\;,
\label{eq:vertexf2gaga}
\end{equation}
where ${\cal{F}}^{\alpha \beta}$ is the electromagnetic field tensor :
\begin{equation}
{\cal{F}}^{\alpha \beta} \left(q, \lambda_\gamma \right)\;=\; 
q^\alpha \, \varepsilon^{\beta *} \left(q, \lambda_\gamma \right) \;-\;
q^\beta \, \varepsilon^{\alpha *} \left(q, \lambda_\gamma \right) \;.
\end{equation}
Using the vertex of Eq.~(\ref{eq:vertexf2gaga}), the
$f_2 \rightarrow \gamma \gamma$ decay width is calculated as :
\begin{equation}
\Gamma({f_2}\,\rightarrow\,\gamma\gamma)\;=\;
\frac{e^4}{80 \pi}\,{g_{{f_2}\gamma\gamma}^2}\,m_{f_2}.
\label{eq:f2gagaw}
\end{equation}
Using the partial width $\Gamma({f_2}\,\rightarrow\,\gamma\gamma)\,=\,
1.32\cdot{10^{-5}}\;\Gamma_0$ \cite{PDG98}, Eq.~(\ref{eq:f2gagaw}) 
determines the value of the coupling constant : 
$g_{{f_2}\gamma\gamma} \simeq$~0.239.
\newline
\indent
Using these couplings and vertices, we can now calculate the 
invariant amplitude for
the process $\gamma\gamma\,\rightarrow\,{f_2}\,\rightarrow\,\pi\pi$ :
\begin{equation}
{\cal{M}}\,(\gamma\gamma\,{\rightarrow^{\!\!\!\!\!\!\!{f_2}}}\,\pi\pi)\;=\;
-i \;2 \, e^2\, \frac{{g_{{f_2}\gamma\gamma}}}{m_{f_2}}\,
{{\cal{F}}^{\mu\delta}}\!(q,\lambda_{\gamma})
{{\cal{F}}_{\delta}}^{\nu}\!(q',\lambda'_{\gamma})\;
{\Delta_{\mu\nu\alpha\beta}}(p,\Lambda)\;
\frac{g_{{f_2}\pi\pi}}{m_{f_2}}\,
{p_\pi^\alpha} \,{{p'_\pi}^{\beta}}\,.
\label{eq:gagapipiampli}
\end{equation}
To determine the $\gamma \gamma \rightarrow \pi \pi$ helicity
amplitudes $F_{\Lambda_\gamma}$ defined in Eq.~(\ref{eq:crossgagapipi}),
we shall evaluate Eq.~(\ref{eq:gagapipiampli}) in the c.m. system. 
For the case of equal photon helicities 
(${\Lambda_\gamma} \,=\,0$) the $f_2$ does not contribute, i.e. 
\begin{equation}
F^{(f_2)}_{\Lambda_\gamma \,=\, 0} \;=\;0 \;.
\end{equation}
For the case of opposite photon helicities (${\Lambda_\gamma}\,=\,2$)
we find after some algebra :
\begin{equation}
F^{(f_2)}_{\Lambda_\gamma \,=\,2} \;=\;
-\frac{e^2}{8}\,
\frac{g_{f_2 \gamma\gamma} \, g_{f_2 \pi\pi}}{m_{f_2}^2}\,
\frac{{t^2}{\beta^2}}{t\,-\,{m_{f_2}^2}\,+\,im_{f_2}{\Gamma_0}}\;
{\sin^2}\theta_{\pi \pi} \;,
\label{eq:f2gagapipilam2}
\end{equation}
where $\theta_{\pi \pi}$ is the pion c.m. angle and $\beta$ the pion
velocity as in Eq.~(\ref{eq:crossgagapipi}).
It is immediately seen from Eq.~(\ref{eq:f2gagapipilam2}) that 
the $f_2$ meson contribution to the d-wave is given by :
\begin{equation}
F^{(f_2)}_{J=2 \; \Lambda_\gamma = 2}(t)\;=\;-\sqrt{\frac{2}{15}}\;
\frac{e^2}{4}\, \frac{g_{f_2 \gamma\gamma} \, g_{f_2 \pi\pi}}{m_{f_2}^2}\,
\frac{t^2 \beta^2}{t\,-\,{m_{f_2}^2}\,+\,im_{f_2}{\Gamma_0}} \;.
\end{equation}
\end{appendix}

\newpage

\begin{table}
\label{tab1}
\caption{The contribution of the dispersion integrals to the spin 
polarizabilities of the proton. The set HDT$(1\pi)$ is calculated from the 
one-pion photoproduction multipoles of the HDT analysis \protect\cite{HDT}, 
while the column HDT gives the total results with the additional contribution 
of inelastic resonance channels. The entries in the last column are the 
predictions of the dispersion calculation of Ref.\protect\cite{lvov98}.}
\begin{center}
\begin{tabular}{c|ddd}
%
$\gamma_{i}-\text{excit.}$ & HDT$(1\pi)$& HDT &Ref.\protect\cite{lvov98} 
\\
\tableline
$\gamma^{(p)}_1$       &+4.83    &+4.33     &+3.1  \\
$\gamma^{(p)}_2$       &-0.81    &-0.74      &-0.8  \\
$\gamma^{(p)}_3$       &-0.30    &-0.02      &+0.3  \\
$\gamma^{(p)}_4$       &+3.19    &+2.93      &+2.7  \\
\tableline
$\gamma^{(p)}_{\rm o}$         &-0.75    &-0.80      &-1.5  \\
$\gamma^{(p)}_{13}$    &+4.23    &+4.29      &+3.7  \\
$\gamma^{(p)}_{14}$    &-1.56    &-1.53       &-2.3  \\
$\gamma^{(p)}_{\pi}$   &+10.41   &+9.46      &+7.8  \\
\end{tabular}
\end{center}
\end{table}

\begin{table}
\label{tab2}
\caption{The same as in Table~I in the case of the neutron.}
\begin{center}
\begin{tabular}{c|ddd}
%
$\gamma_{i}-\text{excit.}$ & HDT$(1\pi)$ &HDT& Ref.\protect\cite{lvov98} 
\\
\tableline
$\gamma^{(n)}_1$         &+7.10    &+7.00       &+6.3   \\
$\gamma^{(n)}_2$         &-0.68    &-0.68       &-0.9   \\
$\gamma^{(n)}_3$         &-1.04    &-0.99       &-0.7   \\
$\gamma^{(n)}_4$         &+3.92    &+3.88       &+3.8   \\
\tableline
$\gamma^{(n)}_{\rm o}$           &-0.06    &-0.09      &-0.4   \\
$\gamma^{(n)}_{13}$      &+5.02    &+5.02      &+4.9   \\
$\gamma^{(n)}_{14}$      &-0.74    &-0.77      &-1.3   \\
$\gamma^{(n)}_{\pi}$     &+14.27   &+14.09     &+13.0  \\
\end{tabular}
\end{center}
\end{table}

\newpage

\begin{figure}[ht]
\vspace{1cm}
\epsfxsize=14cm
\centerline{\epsffile{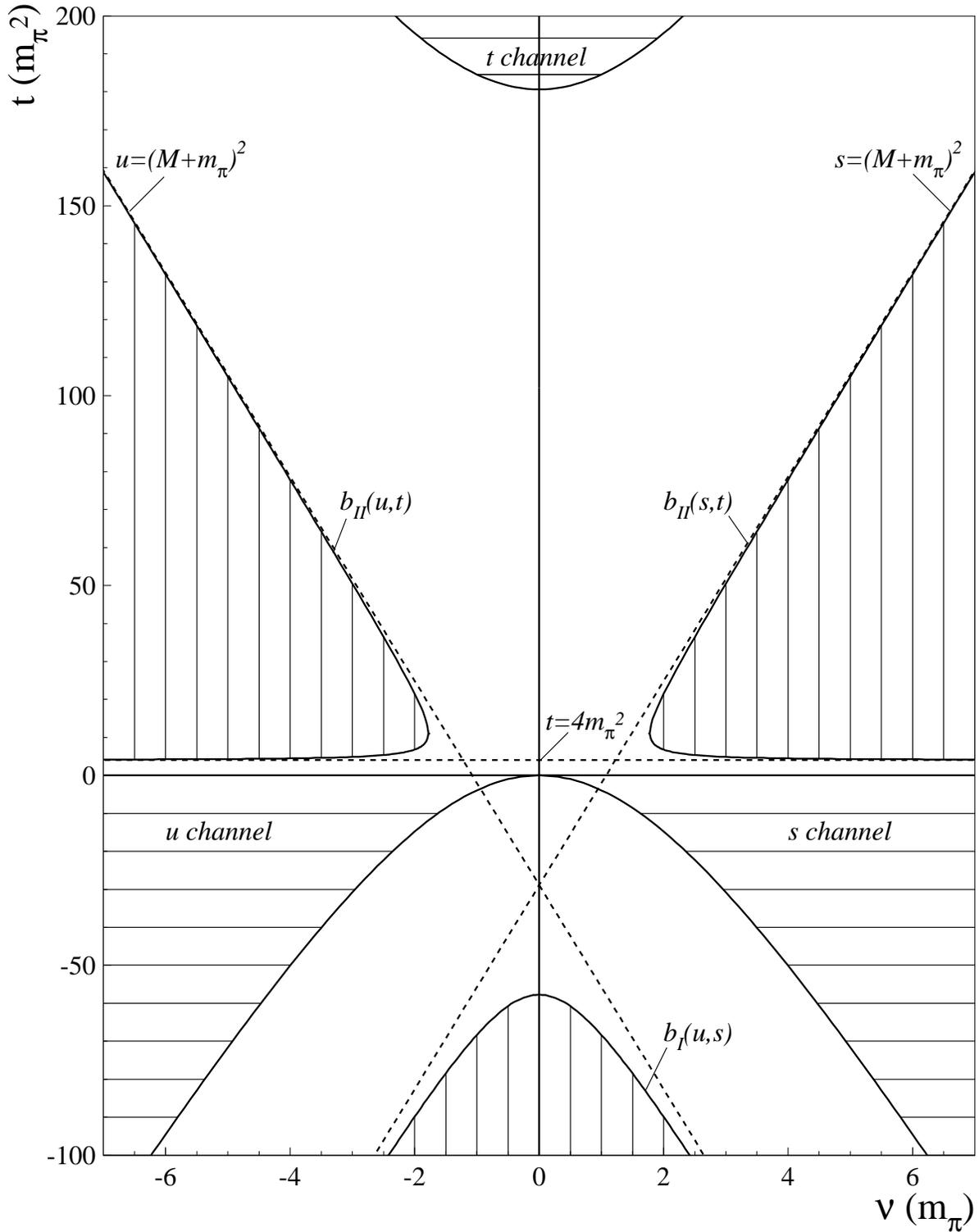}}
\caption[]{The Mandelstam plane for Compton scattering. The physical
  regions are horizontally hatched, whereas the spectral regions are 
  vertically hatched.}
\label{fig:mandelstam}
\end{figure}

\begin{figure}[ht]
\epsfxsize=14 cm
\epsfysize=12. cm
\centerline{\epsffile{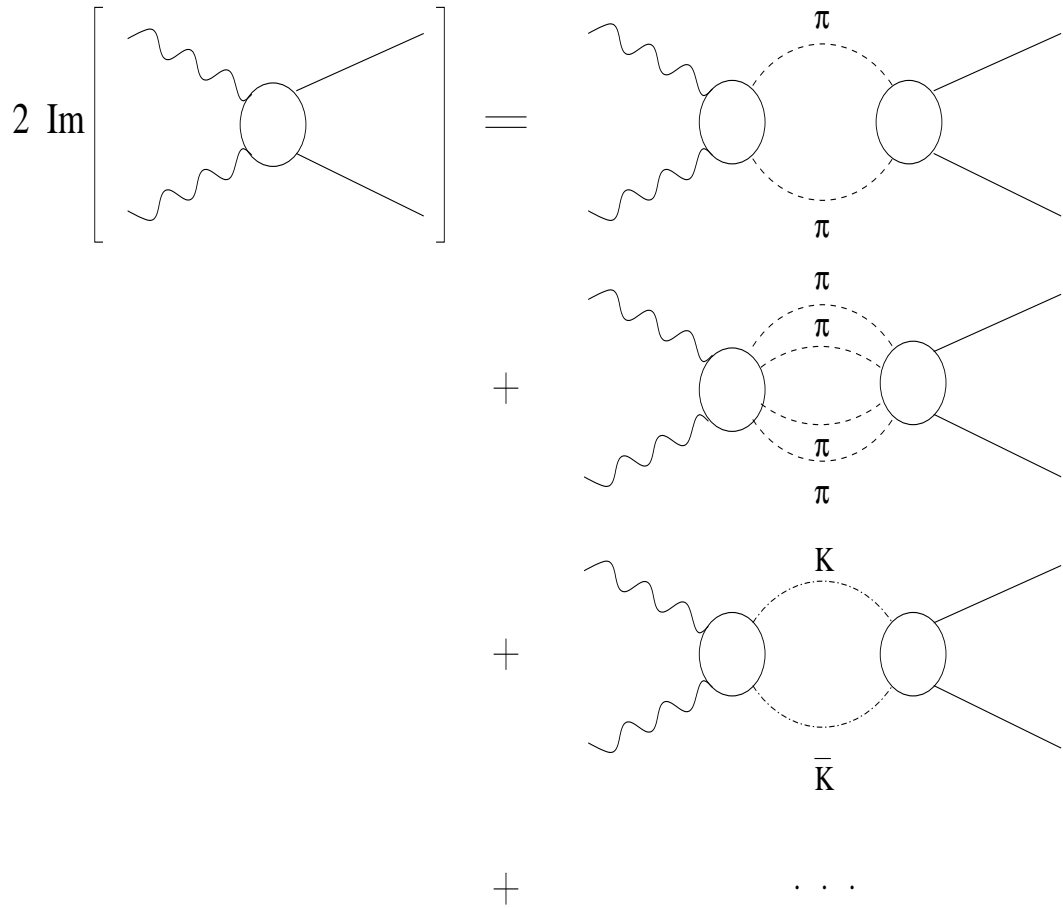}}
\vspace{1.cm}
\caption[]{$t$-channel unitarity diagrams for Compton scattering.}
\label{fig:tunit}
\end{figure}


\begin{figure}[ht]
\epsfysize=17. cm
\centerline{\epsffile{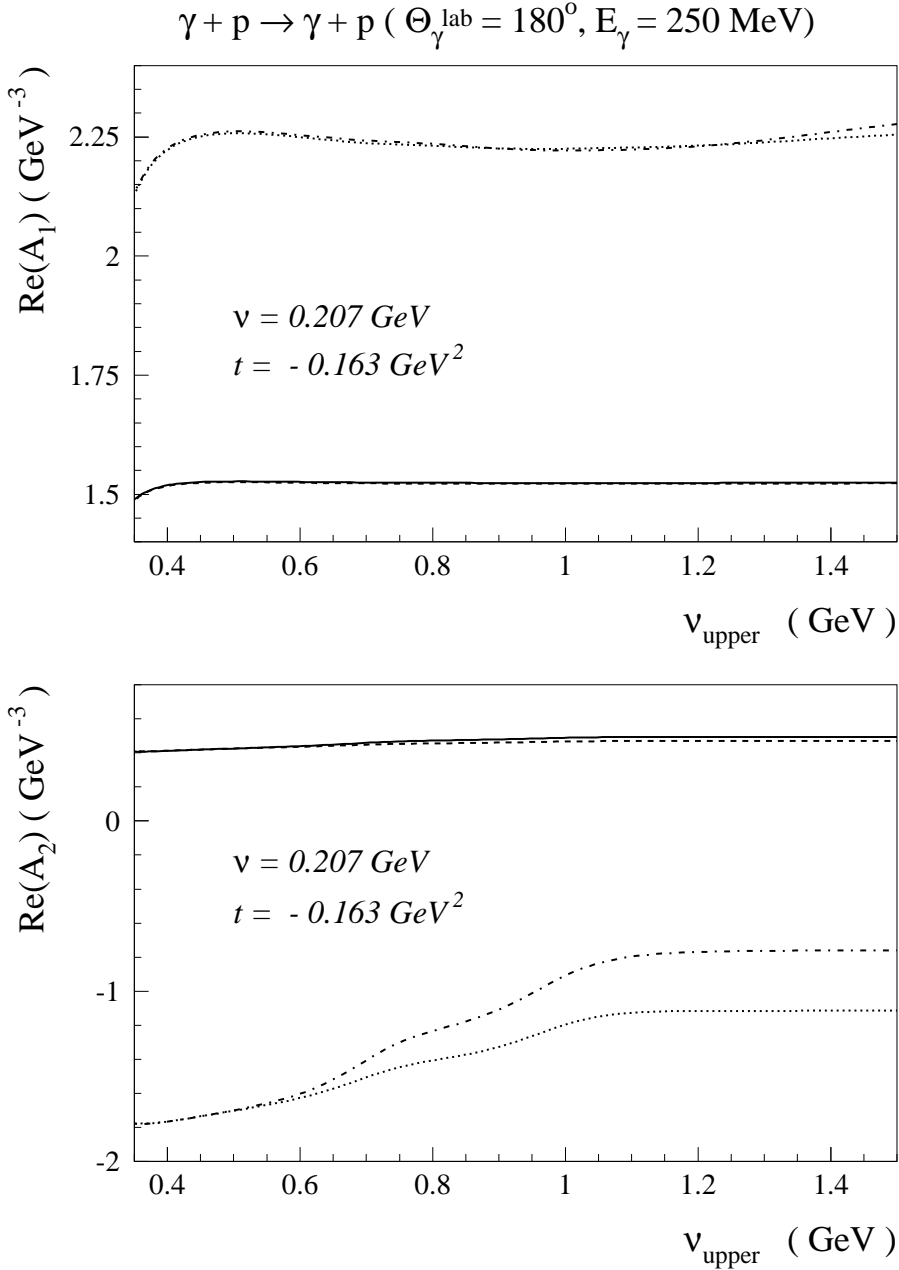}}
\vspace{1.5cm}
\caption[]{Convergence of the $s$-channel integral for the amplitudes 
$A_1$ and $A_2$. 
Results for the unsubtracted dispersion integral of 
Eq.~(\ref{eq:unsub}) for the one-pion channel (dotted lines) and including
the two-pion channel (dashed-dotted lines) in comparison with  
the subtracted dispersion integral of Eq.~(\ref{eq:sub}) 
for the one-pion channel (dashed line) and including the two-pion
channel (full lines), as function of the upper integration limit
$\nu_{\rm upper}$.}
\label{fig:a1a2conv}
\end{figure}

\begin{figure}[ht]
\epsfxsize=13. cm
\centerline{\epsffile{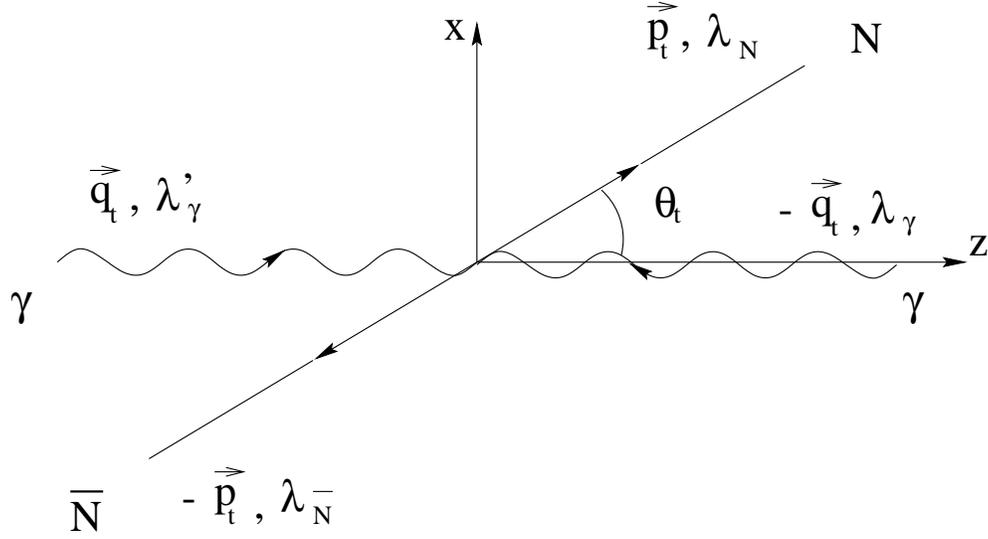}}
\vspace{1.cm}
\caption[]{Kinematics in the c.m. system of the $t$-channel process $\gamma
  \gamma \rightarrow N \bar N$.}
\label{fig:tkinematics}
\end{figure}

\begin{figure}[ht]
\epsfysize=12. cm
\centerline{\epsffile{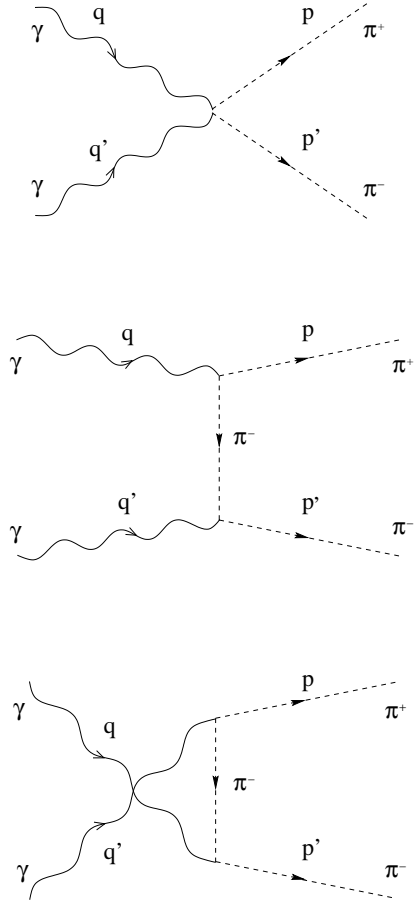}}
\vspace{1.cm}
\caption[]{Born diagrams for the $\gamma \gamma \rightarrow
  \pi^+ \pi^-$ process.}
\label{fig:born}
\end{figure}

\begin{figure}[ht]
\epsfxsize=14 cm
\epsfysize=16. cm
\centerline{\epsffile{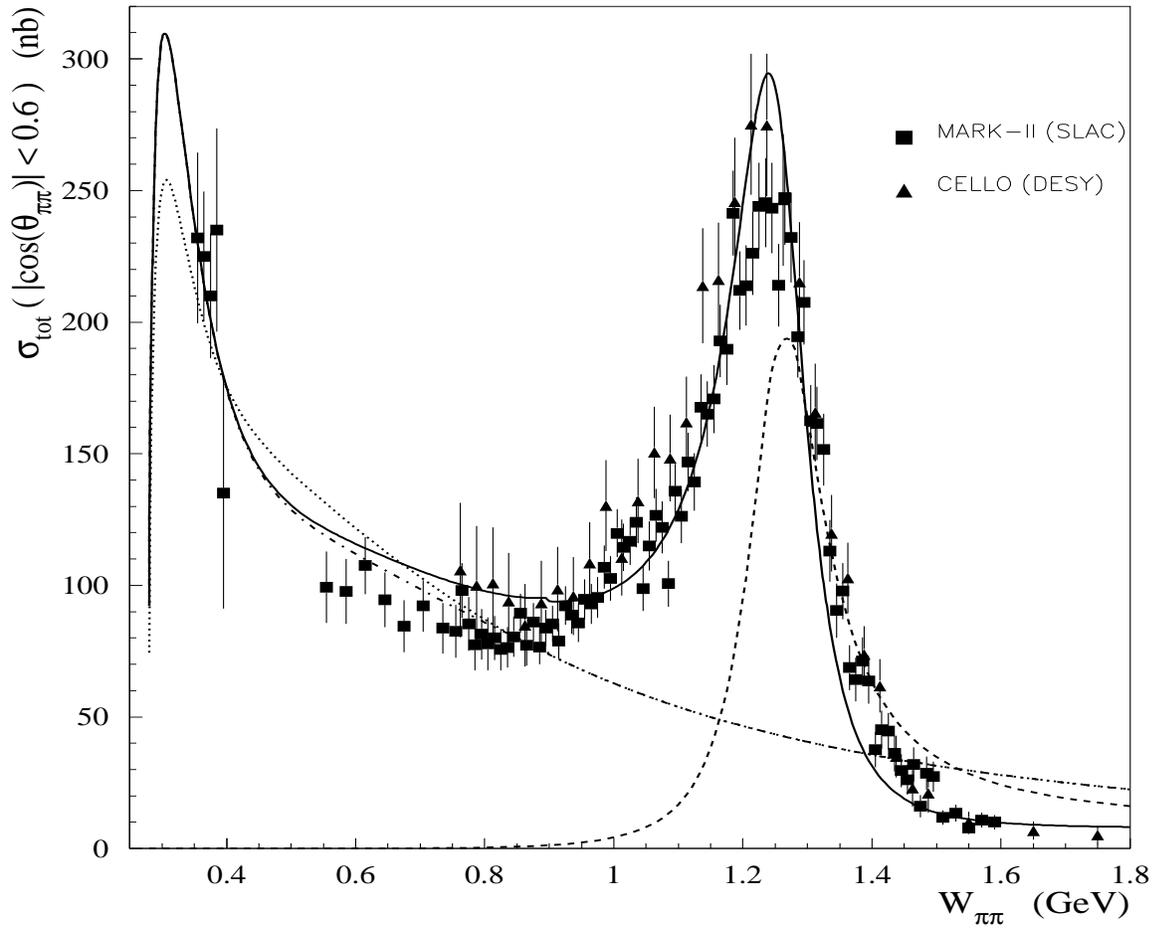}}
\vspace{-1.cm}
\caption[]{Total cross section for the $\gamma \gamma \rightarrow 
\pi^+ \pi^-$ process as function of the c.m. energy :  
Born terms (dotted line), Born amplitude with
  unitarized s-wave (dashed-dotted line), $f_2(1270)$ resonance
  contribution (dashed line) and total amplitude (full line).}
\label{fig:gagapipitot}
\end{figure}

\begin{figure}[ht]
\epsfxsize=14 cm
\epsfysize=18. cm
\centerline{\epsffile{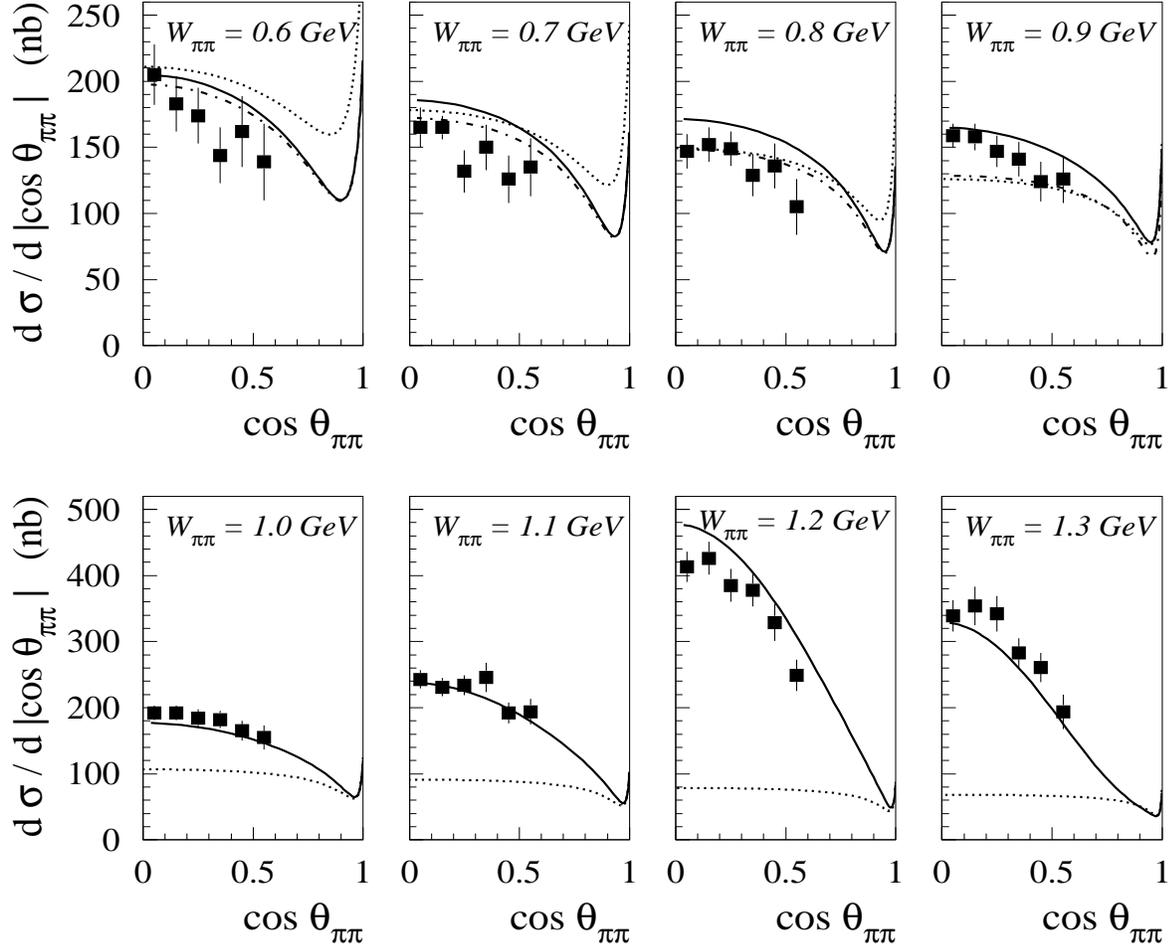}}
\vspace{-2.cm}
\caption[]{Differential cross section at various c.m. energies 
for the $\gamma \gamma \rightarrow \pi^+ \pi^-$ process : 
Born terms (dotted line), Born amplitude with 
unitarized s-wave (dashed-dotted line, only shown at the four 
lower energies), and total amplitude including
the $f_2(1270)$ resonance contribution (full line).}
\label{fig:diffpippim}
\end{figure}

\begin{figure}[ht]
\epsfysize=17. cm
\centerline{\epsffile{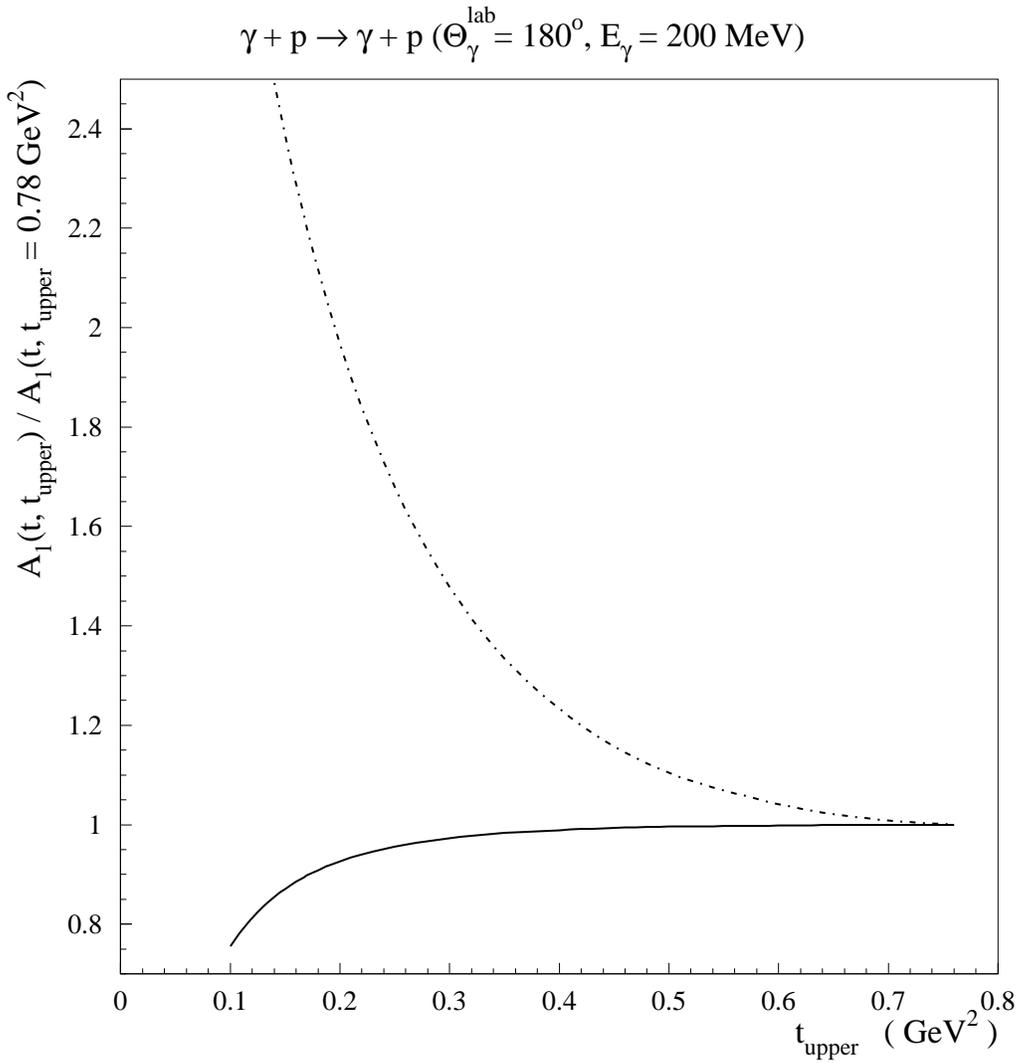}}
\vspace{-1.5cm}
\caption[]{Convergence of the $t$-channel integral for the amplitude 
$A_1$. 
Results for the unsubtracted (dashed curve)
and the subtracted (full curve) $t$-channel dispersion integrals are shown 
as function of the upper integration limit
$t_{\rm upper}$. Both results are normalized to their respective values at
$t_{\rm upper}$ = 0.78 GeV$^2$.}
\label{fig:tchannelconv}
\end{figure}

\begin{figure}[ht]
\epsfxsize=12 cm
\epsfysize=17. cm
\centerline{\epsffile{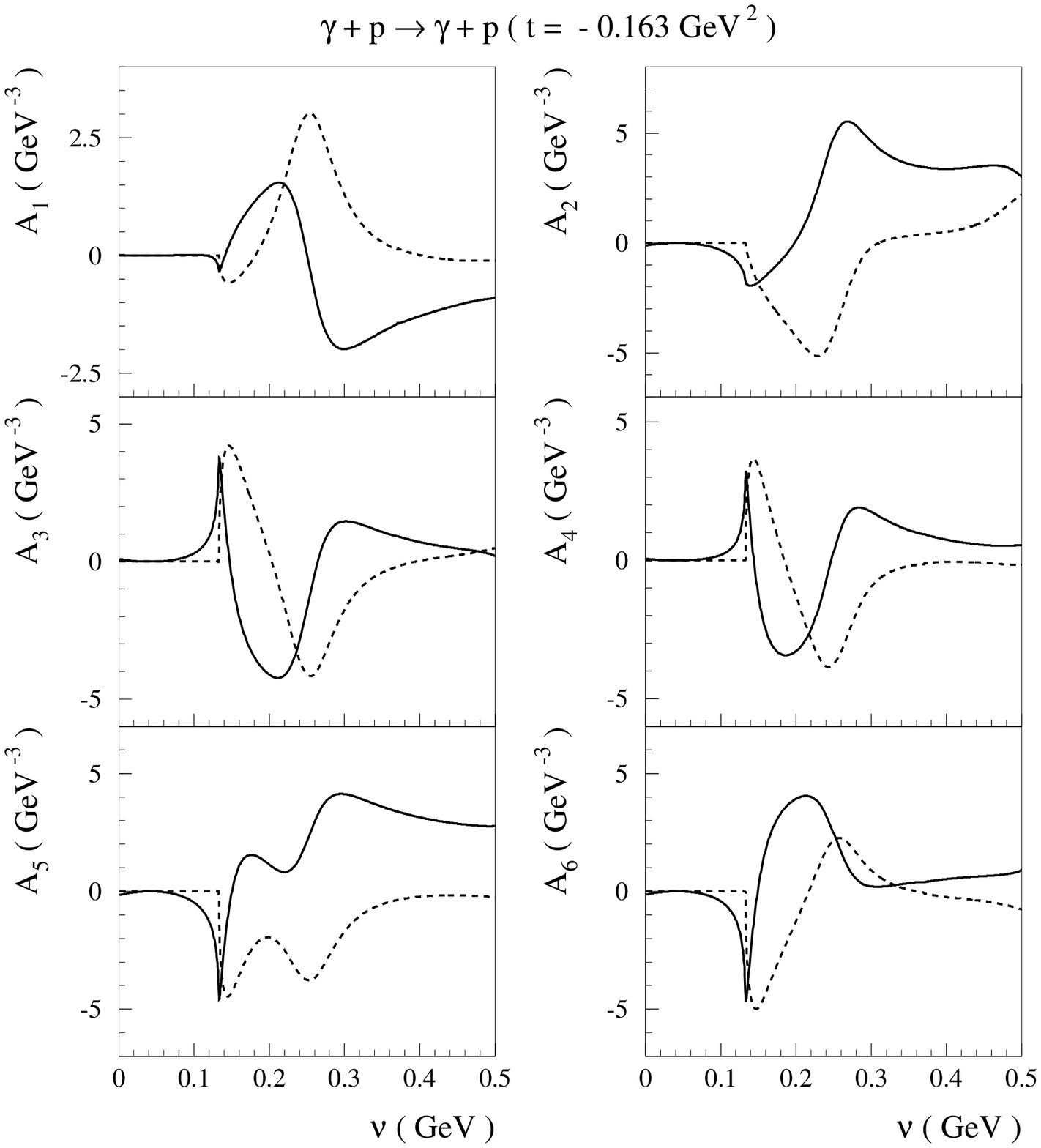}}
\vspace{-.5cm}
\caption[]{Real parts (full lines) 
of the subtracted $s$-channel integral of Eq.~(\ref{eq:sub}) 
and imaginary parts (according to Eq.~(\ref{s-unit})) to 
the invariant Compton amplitudes $A_1,...,A_6$ as function of $\nu$ at
fixed $t$ = -0.163 GeV$^2$.}
\label{fig:a1a6reim}
\end{figure}

\begin{figure}[ht]
\epsfxsize=13 cm
\epsfysize=17. cm
\centerline{\epsffile{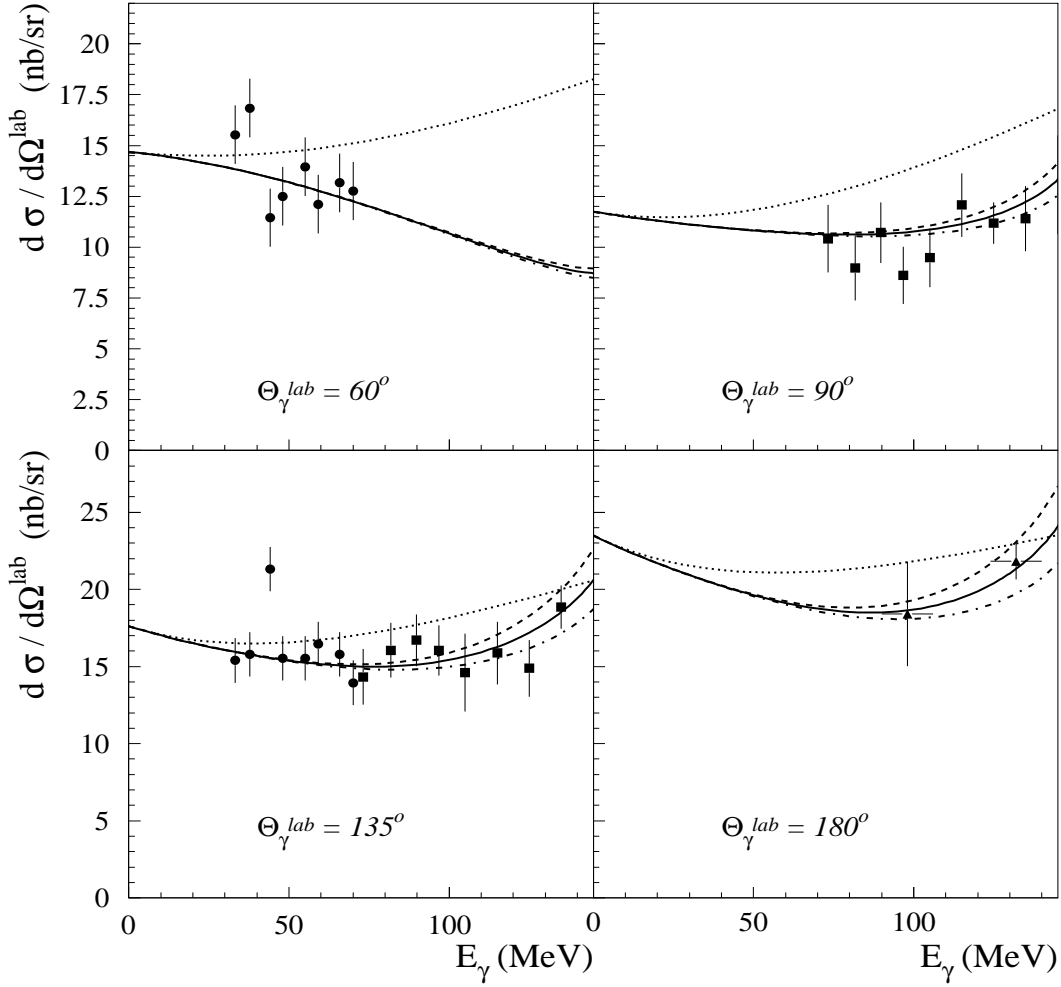}}
\vspace{-1.cm}
\caption[]{Differential cross section for Compton scattering off the proton as 
function of the lab photon energy $E_\gamma$ and at 4 scattering
angles $\Theta_\gamma^{lab}$. 
The Born result is 
given by the dotted lines. The total results of the subtracted
dispersion formalism are shown for fixed $\alpha - \beta = 10$ and 
different values of $\gamma_\pi$ : 
$\gamma_\pi$ = -37 (dashed-dotted lines), 
$\gamma_\pi$ = -32 (full lines) 
and $\gamma_\pi$ = -27 (dashed lines). The data are from 
Ref.~\protect\cite{Federspiel}~(circles), 
Ref.~\protect\cite{Zieger}~(triangles) and 
Ref.~\protect\cite{McGibbon}~(squares).}
\label{fig:threshold_gpi}
\end{figure}

\begin{figure}[ht]
\epsfxsize=13 cm
\epsfysize=17. cm
\centerline{\epsffile{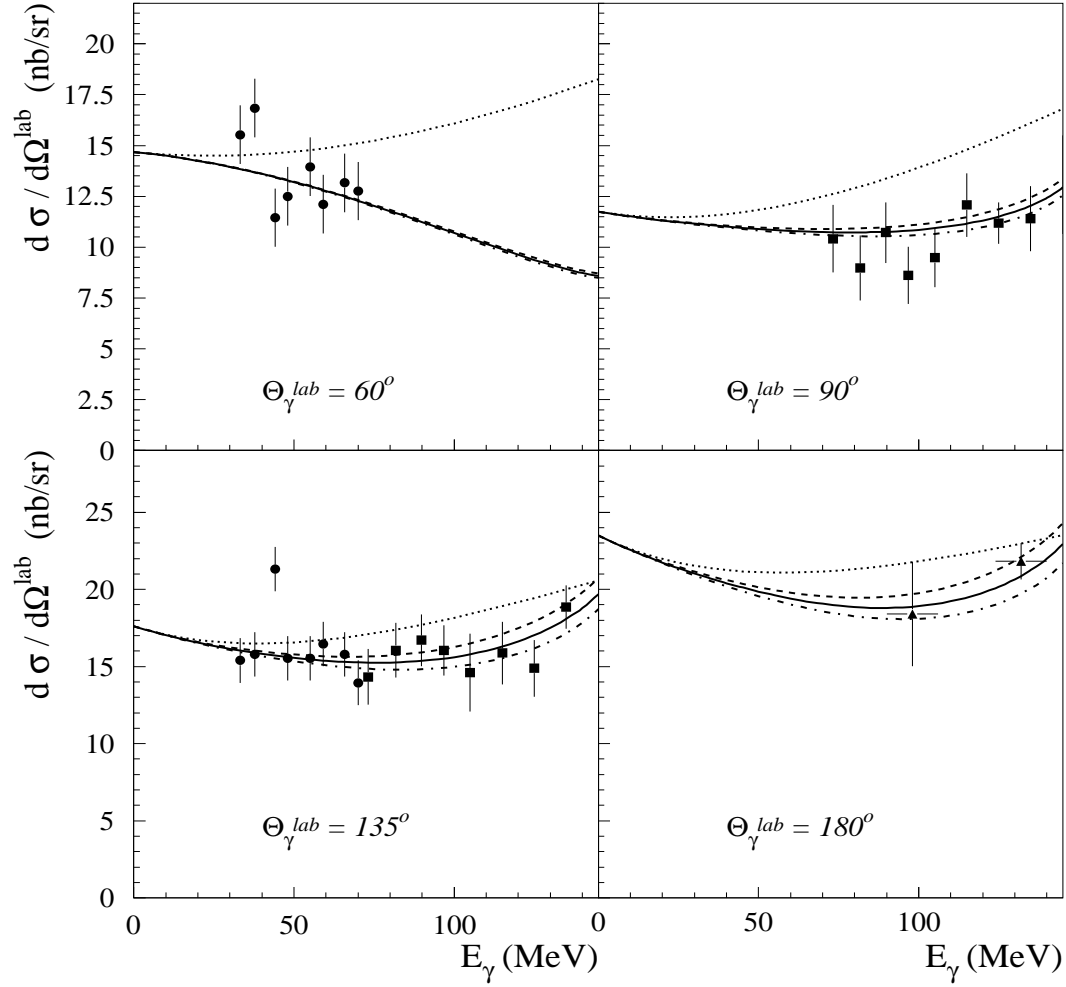}}
\vspace{-1.cm}
\caption[]{Differential cross section for Compton scattering off the proton as 
function of the lab photon energy $E_\gamma$ 
and at 4 scattering angles $\Theta_\gamma^{lab}$ as in 
Fig.~\protect\ref{fig:threshold_gpi}. 
The Born result is given by the dotted lines. 
The total results of the subtracted
dispersion formalism are presented for fixed $\gamma_\pi = -37$ and 
different values of $\alpha - \beta$ : 
$\alpha - \beta$ = 10 (dashed-dotted lines), 
$\alpha - \beta$ = 8 (full lines), 
and $\alpha - \beta$ = 6 (dashed lines). Data as described in 
Fig.~\protect\ref{fig:threshold_gpi}.}
\label{fig:threshold_amb}
\end{figure}

\begin{figure}[ht]
\epsfxsize=13 cm
\epsfysize=17. cm
\centerline{\epsffile{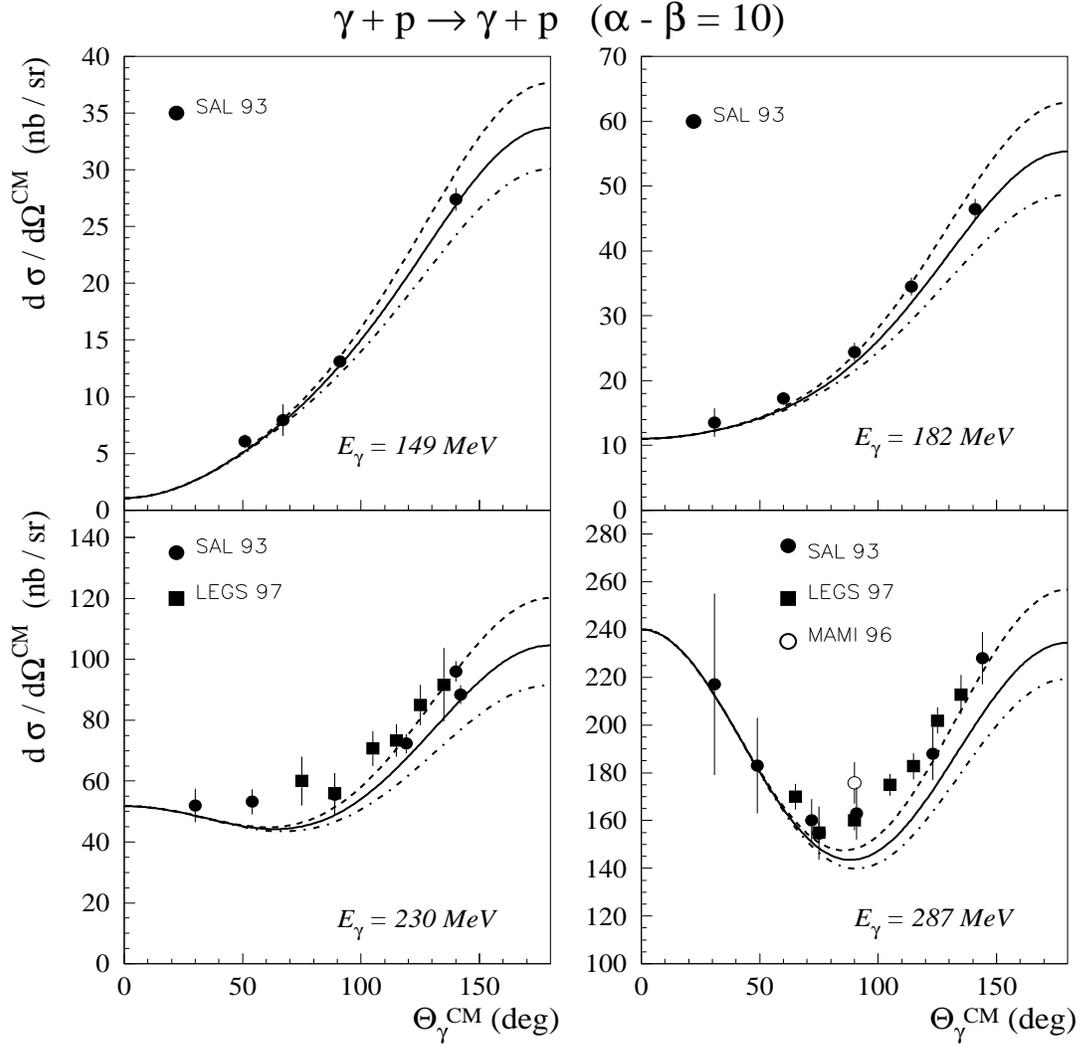}}
\vspace{-1.cm}
\caption[]{Differential cross section for Compton scattering off the proton as 
function of the c.m. photon angle for different lab energies. 
The total results of the subtracted DR formalism 
are presented for fixed $\alpha - \beta = 10$   
and different values of $\gamma_\pi$ : 
$\gamma_\pi$ = -37 (dashed-dotted lines), 
$\gamma_\pi$ = -32 (full lines) 
and $\gamma_\pi$ = -27 (dashed lines). 
The data are from Ref.~\protect\cite{Hallin}~(solid circles), 
Refs.~\protect\cite{Molinari,Peise}~(open circles) and 
Ref.~\protect\cite{Tonnison}~(squares).}
\label{fig:compton_delta_gpi}
\end{figure}

\begin{figure}[ht]
\epsfxsize=13 cm
\epsfysize=17. cm
\centerline{\epsffile{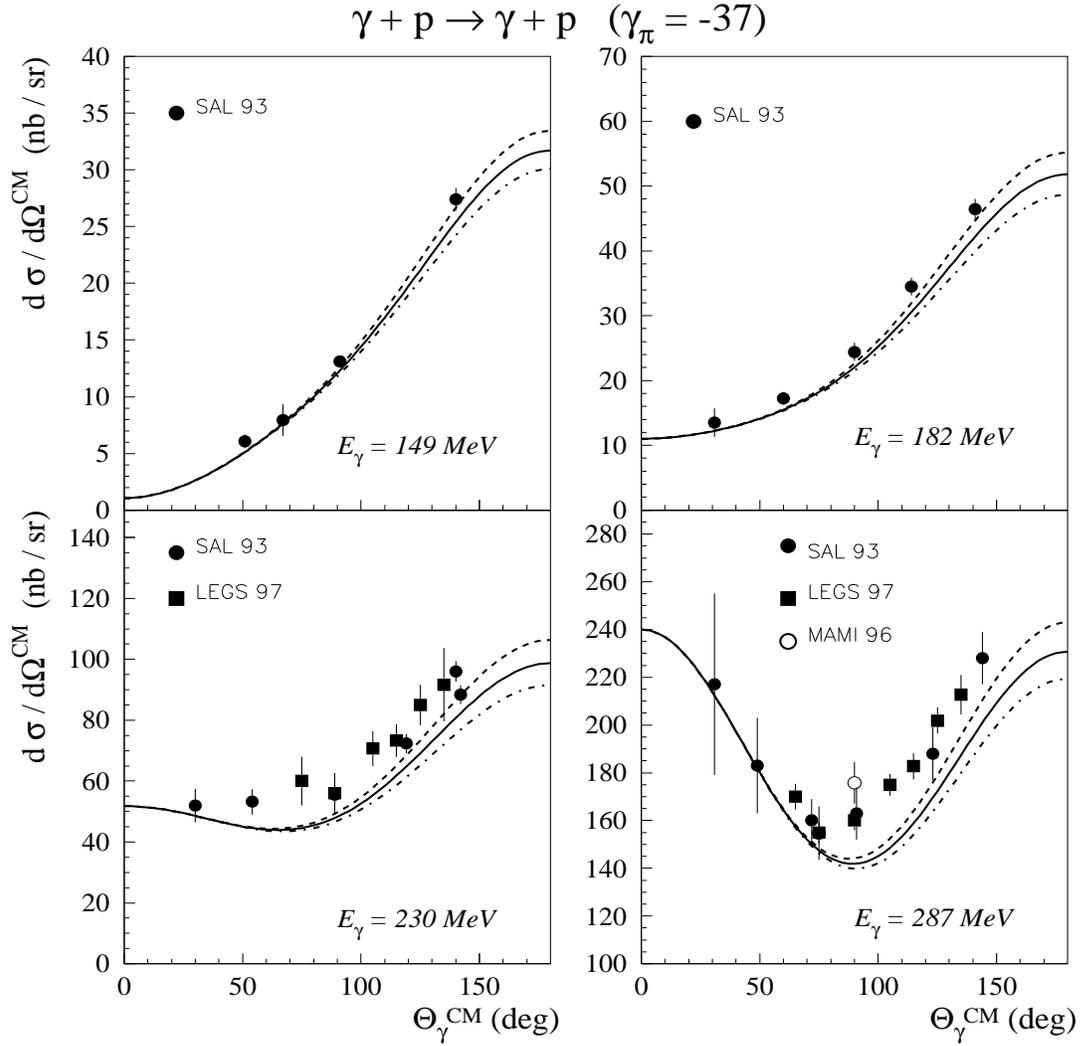}}
\vspace{-1.cm}
\caption[]{Differential cross section for Compton scattering off the proton as 
function of the c.m. photon angle for different lab energies as in 
Fig.~\protect\ref{fig:compton_delta_gpi}. 
The total results of the subtracted DR formalism are presented 
for fixed $\gamma_\pi = -37$ and 
different values of $\alpha - \beta$ : 
$\alpha - \beta$ = 10 (dashed-dotted lines), 
$\alpha - \beta$ = 8 (full lines), 
and $\alpha - \beta$ = 6 (dashed lines). 
Data as described in Fig.~\protect\ref{fig:compton_delta_gpi}.}
\label{fig:compton_delta_amb}
\end{figure}

\begin{figure}[ht]
\epsfxsize=12. cm
\epsfysize=16. cm
\centerline{\epsffile{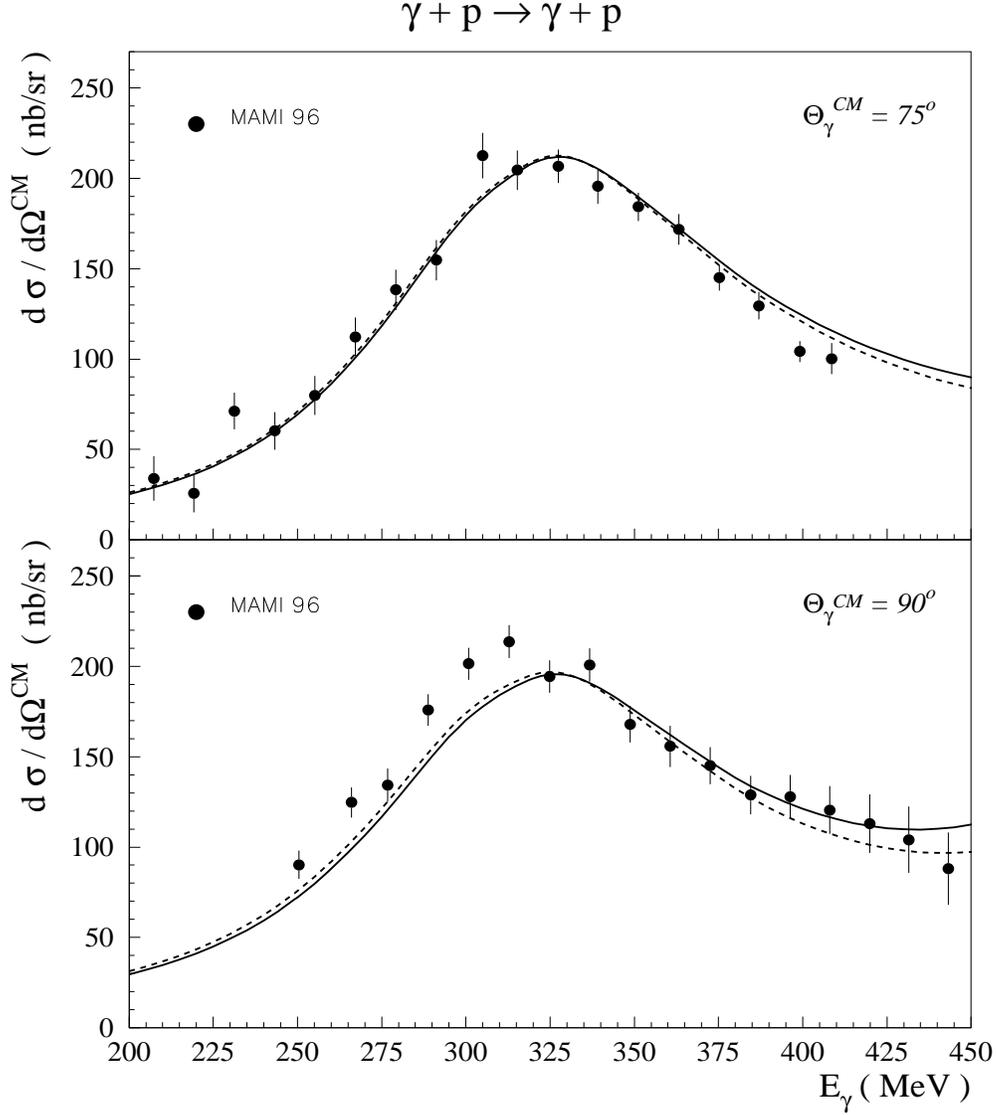}}
\vspace{-.5cm}
\caption[]{Differential cross sections  
for Compton scattering off the proton at fixed c.m. scattering angle 
through the $\Delta$ resonance region. 
The total results of the subtracted DR formalism are shown 
for fixed $\alpha - \beta = 10$ and different values of $\gamma_\pi$ : 
$\gamma_\pi$ = -32 (full lines) 
and $\gamma_\pi$ = -27 (dashed lines). The MAMI data are from 
\protect\cite{Peise} (upper panel) and 
\protect\cite{Molinari} (lower panel).}
\label{fig:fixed_th}
\end{figure}

\begin{figure}[ht]
\epsfxsize=14. cm
\epsfysize=18. cm
\centerline{\epsffile{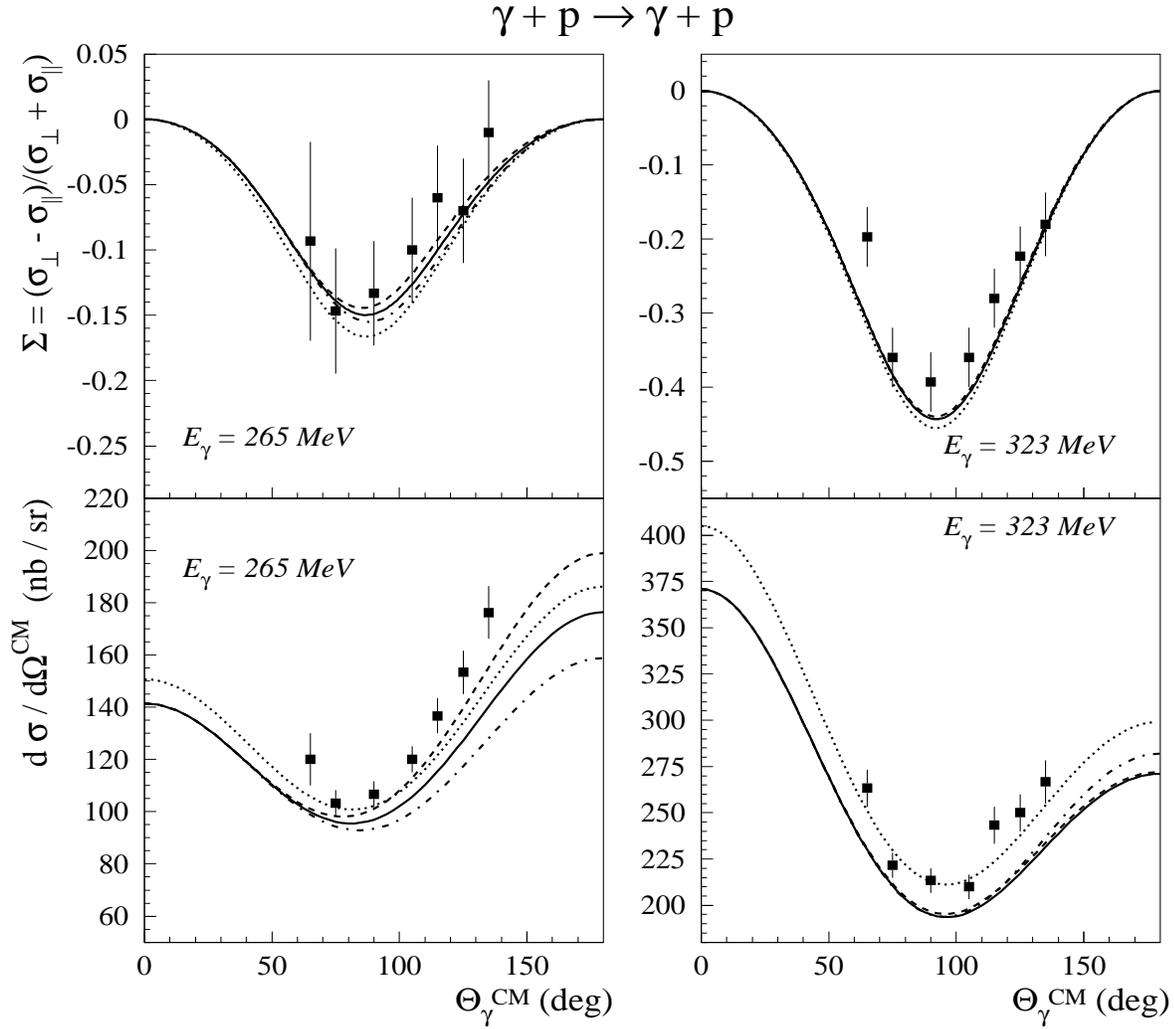}}
\vspace{-1.5cm}
\caption[]{Photon asymmetries (upper panels) 
and differential cross sections (lower panels) 
for Compton scattering off the proton in the $\Delta$ resonance region. 
The total results of the subtracted DR formalism are shown 
for fixed $\alpha - \beta = 10$ and different values of $\gamma_\pi$ : 
$\gamma_\pi$ = -37 (dashed-dotted lines), 
$\gamma_\pi$ = -32 (full lines) 
and $\gamma_\pi$ = -27 (dashed lines). 
We also show the result for $\alpha - \beta$ = 10 
and $\gamma_\pi$ = -32 when increasing the HDT $M_{1+}$ multipole by
2.5 \% (dotted lines). The data are from LEGS 97~\protect\cite{Blanpied97}.}
\label{fig:asymm}
\end{figure}

\begin{figure}[ht]
\epsfxsize=15. cm
\epsfysize=20. cm
\centerline{\epsffile{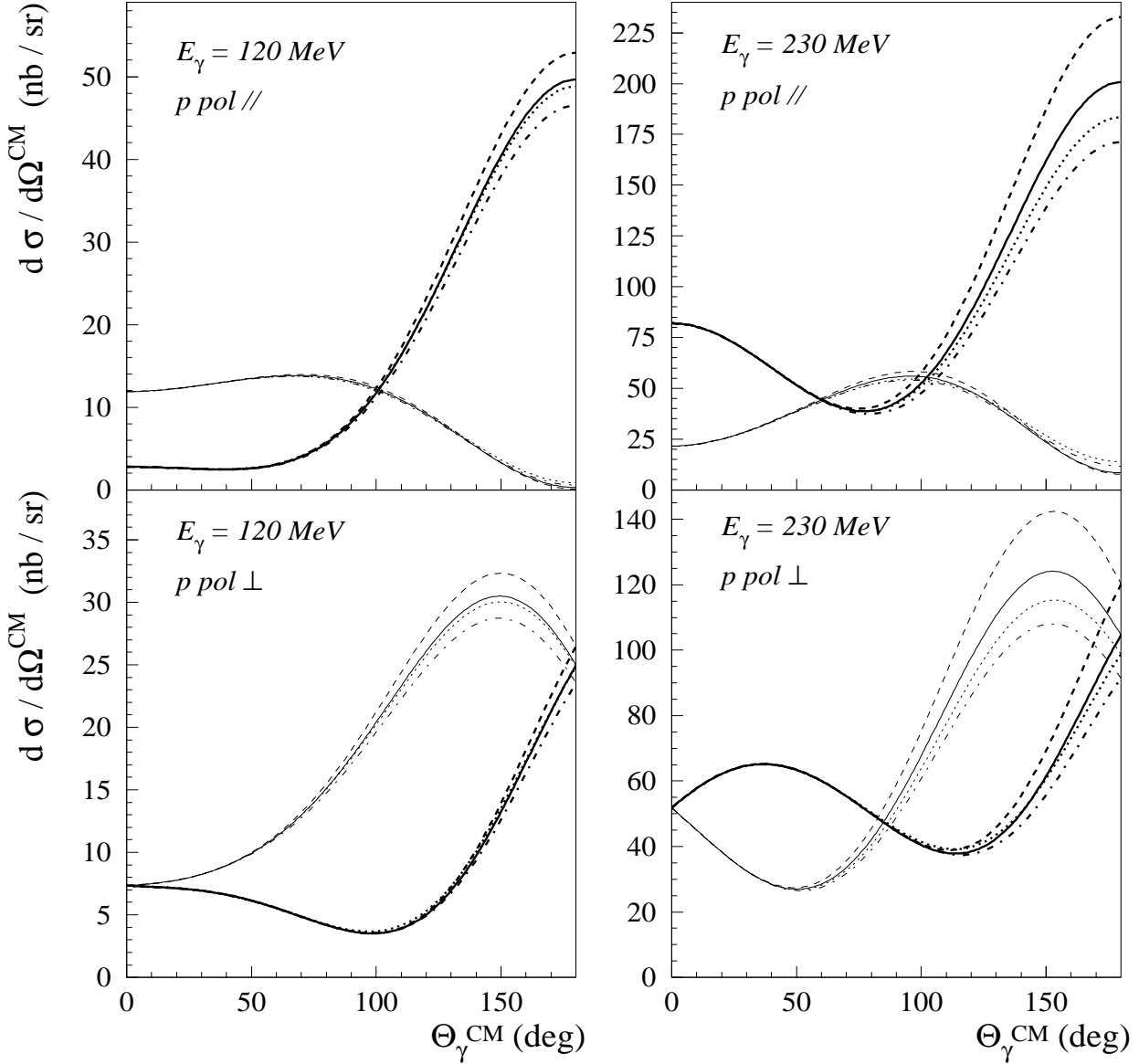}}
\vspace{-1.cm}
\caption[]{Double polarization differential cross sections for Compton 
scattering off the proton, 
with circularly polarized photon and target proton polarized
  along the the photon direction (upper panels) or perpendicular to
  the photon direction and in the plane (lower panels). The thick
  (thin) lines correspond to a proton polarization along the positive
  (negative) direction respectively. 
The results of the dispersion calculation are for 
$\alpha - \beta$ = 10 and different values for $\gamma_\pi$ : 
$\gamma_\pi$ = -32 (full lines), 
$\gamma_\pi$ = -27 (dashed lines) and 
$\gamma_\pi$ = -37 (dashed-dotted lines). 
We also show the result for $\alpha - \beta$ = 8 
and $\gamma_\pi$ = -37 (dotted lines).}
\label{fig:doublepol}
\end{figure}

\end{document}